\documentclass[aps,pra,twocolumn,superscriptaddress]{revtex4-1}
\usepackage{bm}
\usepackage{graphicx}
\usepackage{amssymb,amsmath,amsbsy,amsgen,amsfonts}
\usepackage{dcolumn}
\usepackage{amsthm}
\usepackage{mathrsfs}
\usepackage{latexsym}
\usepackage{array}
\usepackage{amstext}
\usepackage{epsfig}
\usepackage{epstopdf} 
\usepackage{subfigure}
\usepackage{mwe}
\usepackage{color}
\usepackage{units}
\usepackage{calrsfs}
\usepackage{verbatim}
\DeclareMathAlphabet\mathbfcal{OMS}{cmsy}{b}{n}

\newcommand{\be}{\begin{equation}}
\newcommand{\ee}{\end{equation}}
\newcommand{\ba}{\begin{array}}
\newcommand{\ea}{\end{array}}
\newcommand{\bqa}{\begin{eqnarray}}
\newcommand{\eqa}{\end{eqnarray}}

\begin{document}
\title{Robust surface plasmon polaritons on gyrotropic interfaces}

\author{Samaneh Pakniyat}
\email{Pakniyat@uwm.edu}
\address{Department of Electrical Engineering, University of Wisconsin-Milwaukee, 3200 N. Cramer St., Milwaukee, Wisconsin 53211, USA}
\address{Shiraz University of Technology, Shiraz, Fars, Iran}

\author{Alexander M. Holmes}
\email{holmesam@uwm.edu}
\address{Department of Electrical Engineering, University of Wisconsin-Milwaukee, 3200 N. Cramer St., Milwaukee, Wisconsin 53211, USA}

\author{George W. Hanson}
\email{george@uwm.edu}
\address{Department of Electrical Engineering, University of Wisconsin-Milwaukee, 3200 N. Cramer St., Milwaukee, Wisconsin 53211, USA}

\author{S. Ali Hassani Gangaraj}
\address{School of Electrical and Computer Engineering, Cornell University, Ithaca, NY 14853, USA}

\author{Mauro Antezza}
\address{Laboratoire Charles Coulomb (L2C), UMR 5221 CNRS-Universit\'{e} de Montpellier, F-34095 Montpellier, France}
\address{Institut Universitaire de France, 1 rue Descartes, F-75231 Paris Cedex 05, France}

\author{ M\'ario G. Silveirinha}
\address{Instituto Superior T\'{e}cnico, University of Lisbon and Instituto de Telecomunica\c{c}\~{o}es, Torre Norte, Av. Rovisco
	Pais 1, Lisbon 1049-001, Portugal}

\author{ Shahrokh Jam}
\address{Shiraz University of Technology, Shiraz, Fars, Iran}

\author{Francesco Monticone}
\address{School of Electrical and Computer Engineering, Cornell University, Ithaca, NY 14853, USA}

\date{\today}

\begin{abstract}
Unidirectional surface plasmon polaritons (SPPs) at the interface between a gyrotropic medium and a simple medium are studied in a newly-recognized frequency regime wherein the SPPs form narrow, beam-like patterns due to hyperbolic dispersion. The SPP beams are steerable by controlling parameters such as the cyclotron frequency (external bias) or the frequency of operation. The bulk band structure along different propagation directions is examined to ascertain a common bandgap, valid for all propagation directions, which the SPPs cross. The case of a finite-thickness gyrotropic slab is also considered, for which we present the Green function and examine the thickness and loss level required to maintain a unidirectional SPP.
\end{abstract}

\maketitle

\section{Introduction}

Topological surface waves have several important features; namely, they are unidirectional, and they operate in the bulk bandgap of a topologically nontrivial material \cite{18-ozawa2018topological,19-Soljacic2014,23-hasan2010colloquium,25-rechtsman2013photonic,26-chen2014experimental,7-PTI-Notes,17-wang2009observation,casimir}. Upon encountering a discontinuity, they are immune to back-scattering, and because they operate in the bulk bandgap, they do not radiate into the bulk. As such, they are forced to pass over the discontinuity, and the lack of scattering or diffraction makes them interesting from a wave-propagation aspect, and promising for device applications \cite{	13-Ferrite,20-wang2008reflection,21-yu2008one,22-yang2016one}. The topological SPPs can be characterized by an integer invariant (e.g., the Chern number), which cannot change except when the underlying momentum-space topology of the bulk bands is changed \cite{17-wang2009observation,14-Mario-chern,15-Haldane-chern,16-raghu-chern,27-gangaraj2017berry,28-skirlo2014multimode}. Thus, another view of the reflection- and diffraction-free aspect of topological SPPs is that they are governed by the bulk properties so that they are not sensitive to surface features, and can only change qualitatively when the bulk topology changes. A change in topology arises when a bandgap is closed or opened, which occurs for the biased plasma considered here when the bias field is reversed in direction. A static magnetic bias field applied to a plasma breaks time reversal symmetry and leads to topologically non-trivial properties, bringing about the existence of topologically-protected unidirectional photonic surface states \cite{22-yang2016one, 27-gangaraj2017berry, 29-khanikaev2013photonic}.

In this paper, we examine a newly-discovered regime of gyrotropic SPPs \cite{6-PRL}-\cite{9-Trully}, wherein the SPPs are, similar to topological SPPs, unidirectional, operate in a bulk bandgap (and so are diffraction-free), and only change their properties qualitatively when the topology of momentum space is changed. Moreover, they form narrow beam-like patterns, similar to the case of hyperbolic media. 

Unlike in isotropic media, which is described by a single bulk dispersion diagram  identical in every direction, for the anisotropic case, the possibility of a bulk bandgap must be considered in different propagation directions. In this work, we have identified a bulk bandgap common to all propagation directions, within which the SPPs exist. However, it seems difficult or perhaps impossible to assign a topological integer-invariant to describe these SPPs as they propagate in different directions at different frequencies within the gap, and so, strictly-speaking, these SPPs are not topological. Nevertheless, we show that they still exhibit unidirectional propagation and inherent robustness to discontinuities. 

In the following, the common bulk bandgap is discussed, the behavior of the SPPs is determined, and a Green function is obtained for a finite-thickness gyrotropic layer. Additionally, we investigate the back-scattering immune properties of a surface wave propagating at the magnetized plasma-air interface, and also on the surface of a magnetized plasma slab in the presence of a defect in the lower bandgap frequency regime.
\section{Bulk-Mode and SPP Dispersion Analysis}
\label{formulation}
\begin{figure}[!htbp]	\includegraphics[width=0.99\columnwidth]{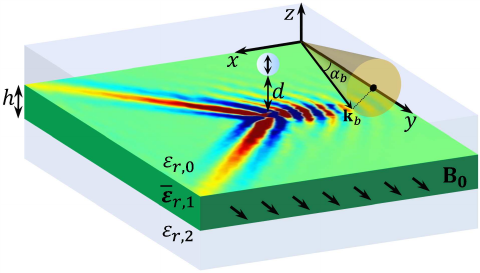}
	\caption{Slab of gyrotropic material with finite thickness, $h$. The slab is biased with a static magnetic field in the xoy plane. A vertical dipole is suspended a distance $d$ above the slab and is responsible for exciting the displayed field pattern near the top surface of the slab. The wavenumber associated with a bulk mode propagating within the slab is denoted $\mathbf{k}_{b}$, and is represented in a local coordinate system where $\alpha_{b}$ denotes the angle which $\mathbf{k}_{b}$ makes with respect to the y-axis.}\label{geom}
\end{figure}

The geometry of interest is depicted in Fig. \ref{geom}, showing a finite-thickness gyrotropic slab immersed in a simple medium characterized by $\varepsilon_{r,0}$ for $z>z_{1}=0$ and $\varepsilon_{r,2}$ for $z<-z_{2}=-h$. The gyrotropic medium is assumed to be a plasma immersed in a static external magnetic field $\mathbf{B}_{0}=\mathbf{\hat{y}}B_{0}$. Assuming time harmonic variation $e^{-j\omega t}$, the magnetized plasma is characterized by the dielectric tensor,
\begin{align}
	\mathbf{\bar{\varepsilon}}_{r}=\varepsilon_{t}\left(  \mathbf{\bar{I}%
	}-\mathbf{\hat{y}\hat{y}}\right)  +j\varepsilon_{g}\left(  \mathbf{\hat{y}%
	}\times\mathbf{\bar{I}}\right)  +\varepsilon_{a}\mathbf{\hat{y}\hat{y},}
	\label{r1}%
\end{align}
where the permittivity elements, $\left\{  \varepsilon_{t},\varepsilon_{a},\varepsilon_{g}\right\}  $ are \cite{coordinateBook}
\begin{align}
	\varepsilon_{t} &  =1-\frac{\omega_{p}^{2}}{\left(  \omega+j\Gamma\right)
	^{2}-\omega_{c}^{2}},\nonumber\\
	\varepsilon_{a} &  =1-\frac{\omega_{p}^{2}}{\omega\left(  \omega
	+j\Gamma\right)  },\ \varepsilon_{g}=\frac{\omega_{c}\omega_{p}^{2}}%
	{\omega\left[  \omega_{c}^{2}-\left(  \omega+j\Gamma\right)  ^{2}\right]
	}.\label{r2}%
\end{align}
such that $\omega_{p}=\sqrt{Nq_{e}^{2}/m_{e}\varepsilon_{0}}$, $\omega_{c}=-q_{e}B_{0}/m$, and $\Gamma=1/\tau$ denote the plasma, cyclotron, and collision frequencies, respectively, where $N$ is the free electron density, $q_{e}=-e$ is the electron charge, $m_{e}$ is the electron mass, and $\tau$ is the relaxation time between collisions. The above model is local; as studied in \cite{Shi}, a nonlocal Drude model leads to the presence of a backward propagating modes. However, the effect of non-locality is evident only for very large wavenumbers and the backward waves vanish when considering realistic levels of loss \cite{9-Trully}, and so non-locality is ignored here. 

\subsection{Dispersion of bulk modes in a gyrotropic medium -- the existence of a common bandgap}
The characteristics of the bulk modes in an anisotropic medium depend on the direction of propagation. In a structure exhibiting bulk band-gaps, these will also be direction-dependent. In this section, we study the bulk dispersion behavior of a gyrotropic medium in order to identify a bulk bandgap, common to all propagation directions. We begin with a plane wave having wave vector, $\mathbf{k}_{b}$, propagating in a gyrotropic medium at angle, $\alpha_{b}$, with respect to the bias field ($y$) direction. Assuming a plane wave solution to Maxwell's equations leads to a homogeneous system of equations for which non-trivial solutions are obtained when \cite{coordinateBook}
\begin{equation}
	\left\vert k_{0}^{2}\mathbf{\bar{\varepsilon}}_{r}-k_{b}^{2}\mathbf{\bar{I}%
	}+\mathbf{k}_{b}\mathbf{k}_{b}\right\vert =0, \label{r3}%
\end{equation}
where $\mathbf{k}_{b}=\mathbf{k}_{t}+\mathbf{\hat{y}}k_{y}$ such that $\left\vert\mathbf{k}_{t}\right\vert=k_{b}\sin  \alpha_{b}$ and $k_{y}=k_{b}\cos \alpha_{b}$. Evaluation of the determinant leads to the dispersion equation for the bulk modes,
\begin{align}
	0 &  =k_{b}^{2}k_{0}^{2}\left\{  \left[  \varepsilon_{t}\left(  \varepsilon
	_{t}+\varepsilon_{a}\right)  -\varepsilon_{g}^{2}\right]  \sin^{2}\alpha
	_{b}+2\varepsilon_{t}\varepsilon_{a}\cos^{2}\alpha_{b}\right\}  \nonumber\\
	&  -k_{b}^{4}\left[  \varepsilon_{t}\sin\alpha_{b}+\varepsilon_{a}\cos
	^{2}\alpha_{b}\right]  -k_{0}^{4}\left(  \varepsilon_{t}^{2}-\varepsilon
	_{g}^{2}\right)  \varepsilon_{a}.\label{dispersion_eq}%
\end{align}
\begin{figure}[!tbp]
	\includegraphics[width=0.99\columnwidth]{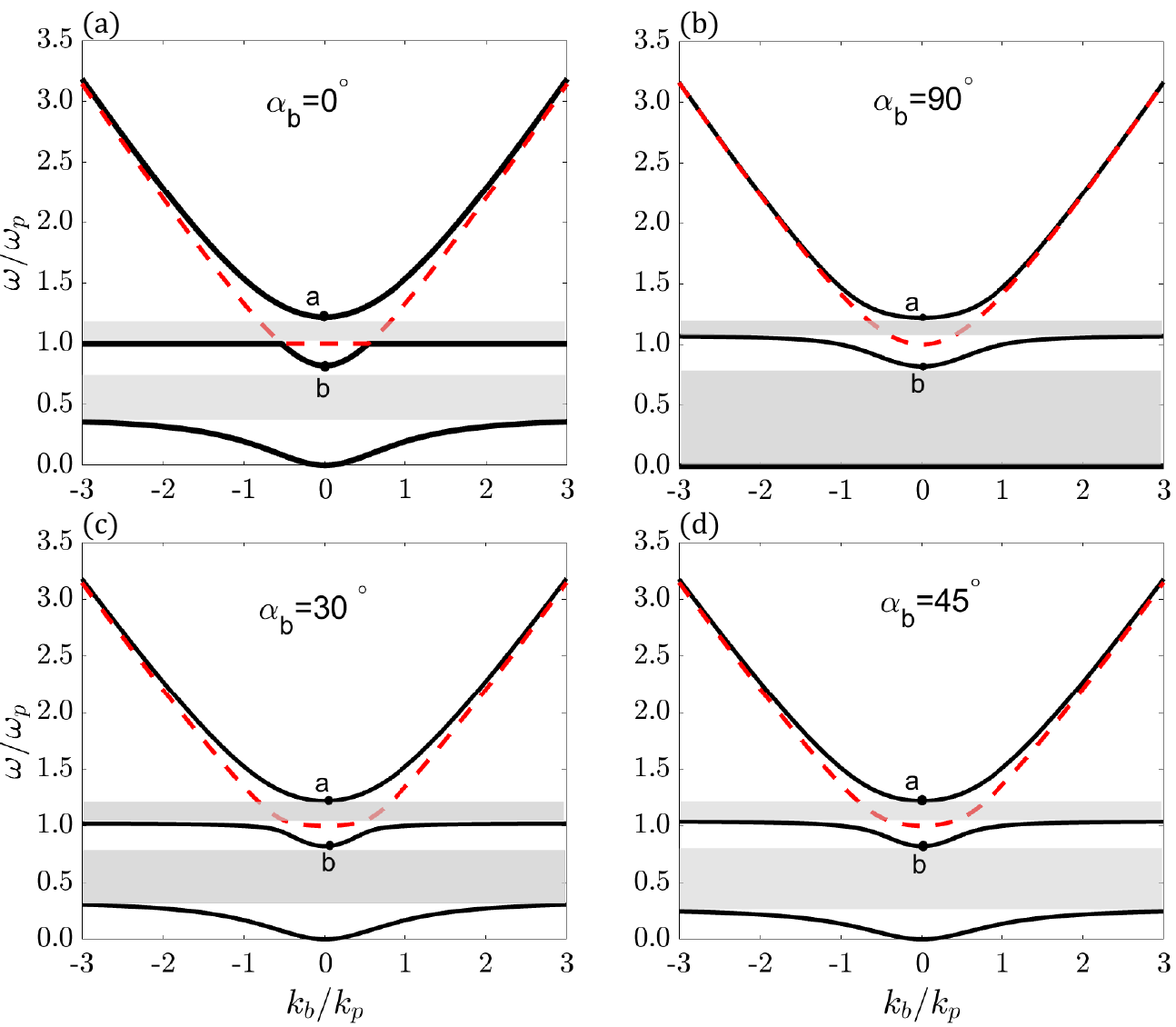}
	\caption{Dispersion diagram of plasma bulk modes for different angles of propagation, where $k_p=\omega_p /c$. Gray shaded regions highlight bandgaps in the dispersion. The dashed red line corresponds to an ordinary wave (independent of bias) while the solid black lines correspond to the extraordinary wave (dependent on bias).}
	\label{bulk}
\end{figure}
The dispersion diagrams associated with the bulk modes of a magneto-plasma are shown in Fig. \ref{bulk}. We consider $\omega _{p}=2\pi$(20 THz) and $\omega_{c}/\omega _{p}=0.4$ here and throughout the rest of the paper. Figures \ref{bulk}a and \ref{bulk}b show the dispersion of bulk modes which propagate parallel ($\alpha_{b}=0^{\circ}$) and perpendicular ($\alpha_{b}=90^{\circ}$) to the magnetic bias, respectively. In the parallel case, the two intersection points correspond to Weyl points arise from crossings between longitudinal plasma modes and transverse helical modes \cite{10-weyl}. Figures \ref{bulk}c and \ref{bulk}d show the dispersion for two arbitrary angles in the range, $0^{\circ} < \alpha_{b} < 90^{\circ}$. As seen in Fig. \ref{bulk}, there are four branches of the dispersion. The second branch from the top (dashed red) corresponds to an ordinary wave, independent of the magnetic bias, which does not lead to a topological SPP. Two bandgaps form between the other three branches as shown in the shaded regions of Fig. \ref{bulk}. The size of the bandgaps depend on the propagation direction as well as the magnetic bias field strength. The upper bandgap is smallest when $\alpha_{b}=90^{\circ}$. Conversely, the lower band-gap is smallest when $\alpha_{b}=0^{\circ}$. As such, we take the smallest upper (lower) band-gap to represent the common upper (lower) bandgap for all propagation angles, $0^{\circ}<\alpha_{b}<90^{\circ}$. Points $a$ and $b$ do not change with the propagation angle. The common bandgap and its impact on surface waves is considered further in the following.

\subsection{Surface Plasmon Polariton Dispersion}\label{using}

A surface wave that propagates along the interface between a gyrotropic medium and an isotropic medium has a longitudinal wave vector component, $\mathbf{k}_{s}=\mathbf{\hat{x}}k_{x}+\mathbf{\hat{y}}k_{y}$, where the propagation angle, $\phi_{s}$, is made with respect to the $x$ axis. Solving the bulk dispersion equation (\ref{dispersion_eq}), we obtain $\mathbf{k}_{b,i}=\mathbf{\hat{x}}k_{x}+\mathbf{\hat{y}}k_{y}+\mathbf{\hat{z}}mk_{z,i}$ for $i\in\left\{1,2\right\}$ and 
$m\in\left\{\pm\right\}$ where we define $k_{z,i}= j\gamma_{i}$ such that \cite{2-silveirinha2017topological}
\begin{align}
	\gamma_{i}^{2}  &  =k_{x}^{2}\mp\frac{1}{2\varepsilon_{t}}\sqrt{\kappa
	}\nonumber\\
	&  -\frac{1}{2\varepsilon_{t}}\left[  \left(  \varepsilon_{t}\left(
	\varepsilon_{t}+\varepsilon_{a}\right)  -\varepsilon_{g}^{2}\right)  k_{0}%
	^{2}-\left(  \varepsilon_{a}+\varepsilon_{t}\right)  k_{y}^{2}\right],
	\label{r5}%
\end{align}
and
\begin{align}
	\kappa &  =\left[  \left(  \varepsilon_{t}\left(  \varepsilon
	_{t}+\varepsilon_{a}\right)  -\varepsilon_{g}^{2}\right)  k_{0}^{2}-\left(
	\varepsilon_{a}+\varepsilon_{t}\right)  k_{y}^{2}\right]  ^{2}\nonumber\\
	&  -4\varepsilon_{t}\varepsilon_{a}\left[  \left(  \varepsilon_{g}%
	+\varepsilon_{t}\right)  k_{0}^{2}-k_{y}^{2}\right]  \left[  \left(
	\varepsilon_{t}-\varepsilon_{g}\right)  k_{0}^{2}-k_{y}^{2}\right].
\label{r6}%
\end{align}

The dispersion relation for the SPP can be obtained by matching the tangential components of the electric and magnetic fields at the interface [Appendix C, \cite{3-PRA}], leading to the $4\times4$ system of homogeneous equations
\begin{equation}
	\left(
	\begin{array}
	[c]{cccc}%
	\beta_{1}^{-} & \beta_{2}^{-} & k_{y} & j\gamma k_{x}\\
	k_{y}\theta_{1} & k_{y}\theta_{2} & -k_{x} & j\gamma k_{y}\\
	k_{y}\phi_{1}^{-} & k_{y}\phi_{2}^{-} & j\gamma k_{x} & -k_{y}k^{2}\\
	-\delta_{1}k_{t,1}^{2} & -\delta_{2}k_{t,2}^{2} & j\gamma k_{y} & k_{x}k^{2}%
	\end{array}
	\right)  \left(
	\begin{array}
	[c]{c}%
	A_{1}^{-}\\
	A_{2}^{-}\\
	B_{1}^{+}\\
	B_{2}^{+}%
	\end{array}
	\right)  =\mathbf{0,}\label{r7}%
\end{equation}
where $k^{2}=k_{0}^{2}\varepsilon_{r,0}$, $\gamma=\sqrt{k_{x}^{2}+k_{y}^{2}-k^{2}}$, and
\begin{align}
	\delta_{i} &  =j\varepsilon_{g}/\xi_{i},\ \theta_{i}=-k_{t,i}^{2}/\varpi
	_{i},\nonumber\\
	\beta_{i}^{m} &  =k_{x}-mk_{z,i}\delta_{i},\ \phi_{i}^{m}=\delta_{i}%
	k_{x}-mk_{z,i}\left(  \theta_{i}-1\right),
\end{align}
such that $\xi_{i}=k_{0}^{2}\varepsilon_{t}-k_{i}^{2}$ and $\varpi_{i}=k_{0}^{2}\varepsilon_{a}-k_{t,i}^{2}$. Non-trivial solutions are obtained when the determinant of the coefficient matrix on the left hand side of (\ref{r7}) is set equal to zero. Evaluation of the determinant and division through by $-jk_{s}^{2}k_{y}/\varpi_{1}\varpi_{2}\xi_{1}\xi_{2}\neq0$, leads to
\begin{align}
	0 &  =\left(  k_{y}^{2}+k_{z}^{2}\right)  n_{A}-k_{x}n_{B}^{-}+k_{x}k_{y}%
	^{2}n_{C}^{-}\nonumber\\
	&  -\left(  k_{x}^{2}+k_{z}^{2}\right)  n_{D}^{-}-jk_{z}\left(  n_{E}%
	^{-}-\varepsilon_{r,0}\chi^{-}\right),
	\label{SPPdisp}
\end{align}
where $k_{z}=j\gamma$ and the quantities $n_{A}$, $n_{B}^{-}$, $n_{C}^{-}$, $n_{D}^{-}$ and $n_{E}^{-}$ are defined in the Appendix.
\begin{figure}[tbh]
	\includegraphics[width=0.99\columnwidth]{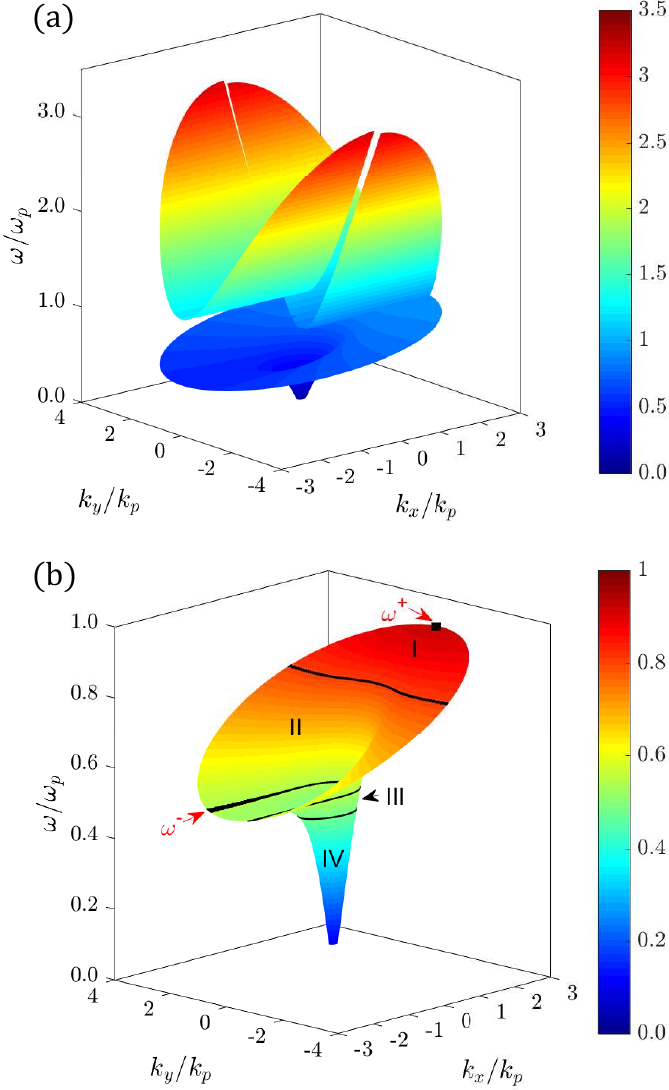}
	\caption{SPP dispersion surface for a biased-plasma-vacuum interface, obtained by solving for the roots of (\protect\ref{SPPdisp}), for $\omega_{c}=0.4\omega_{p}$. (a) Perspective (zoomed) view of the upper and lower bands. (b) Perspective view of the lower band where the solid black lines are the equi-frequency contours for a few representative frequencies and $\omega^{\pm}$ outline the region of SPP resonance. The designations, I-IV, refer to Fig. \ref{commonBG}.}
	\label{3DSPP}
\end{figure}
In what follows, we assume that the upper medium is characterized by $\varepsilon_{r,0}=1$. For the well-studied \cite{4-davoyan2013theory} case of propagation perpendicular to the bias ($k_{y}=0$) the SPP dispersion is found to be
\begin{equation}
	\sqrt{k_{x}^{2}-k_{0}^{2}}+\frac{\sqrt{k_{x}^{2}-k_{0}^{2}\varepsilon_{eff}}%
	}{\varepsilon_{eff}}=\frac{\varepsilon_{g}k_{x}}{\varepsilon_{t}%
	\varepsilon_{eff}}, \label{r10}%
\end{equation}
where $\varepsilon_{eff}=\left(  \varepsilon_{t}^{2}-\varepsilon_{g}^{2}\right)  /\varepsilon_{t}$. For $k_{y}\neq0$, the general dispersion equation (\ref{SPPdisp}) must be used.

As considered in recent photonic topological work \cite{5-PRA-june}, we are interested in bulk-bandgap crossing SPPs. Since the upper bandgap for the perpendicular case and lower bandgap for the parallel case determine the common bandgap of all bulk modes, we consider the SPP modes that cross these two common bandgaps.

A surface mode propagating in the xoy plane generally possesses two wave vector components, $k_{x}$ and $k_{y}$. Therefore, a three-dimensional surface is needed to completely describe the SPP dispersion. As shown in Fig. \ref{3DSPP} the SPP modes form two frequency bands. The upper band is asymmetric about the $k_{x}=0$ plane and symmetric about the $k_{y}=0$ plane and passes through the upper bulk bandgap. The upper band of SPP modes in the magnetized plasma-opaque structure lead to topological unidirectional and back-scattering immune SPPs which has been well studied in \cite{7-PTI-Notes, 6-PRL, 5-PRA-june, 8-three-defect}. For the case that the magnetized plasma immersed in a transparent medium, the upper band represents fast surface waves. These surface waves leak rapidly into the transparent medium. Similarly, the lower band is asymmetric about the $k_{x}=0$ plane and symmetric about the $k_{y}=0$ plane. Furthermore, this lower band passes through the lower bulk bandgap. Dispersion in this lower band leads to beam-like SPPs and has only recently been considered in our previous paper \cite{9-Trully}; this is the main subject of this work.
%
\begin{figure}[!b]
	\includegraphics[width=0.99\columnwidth]{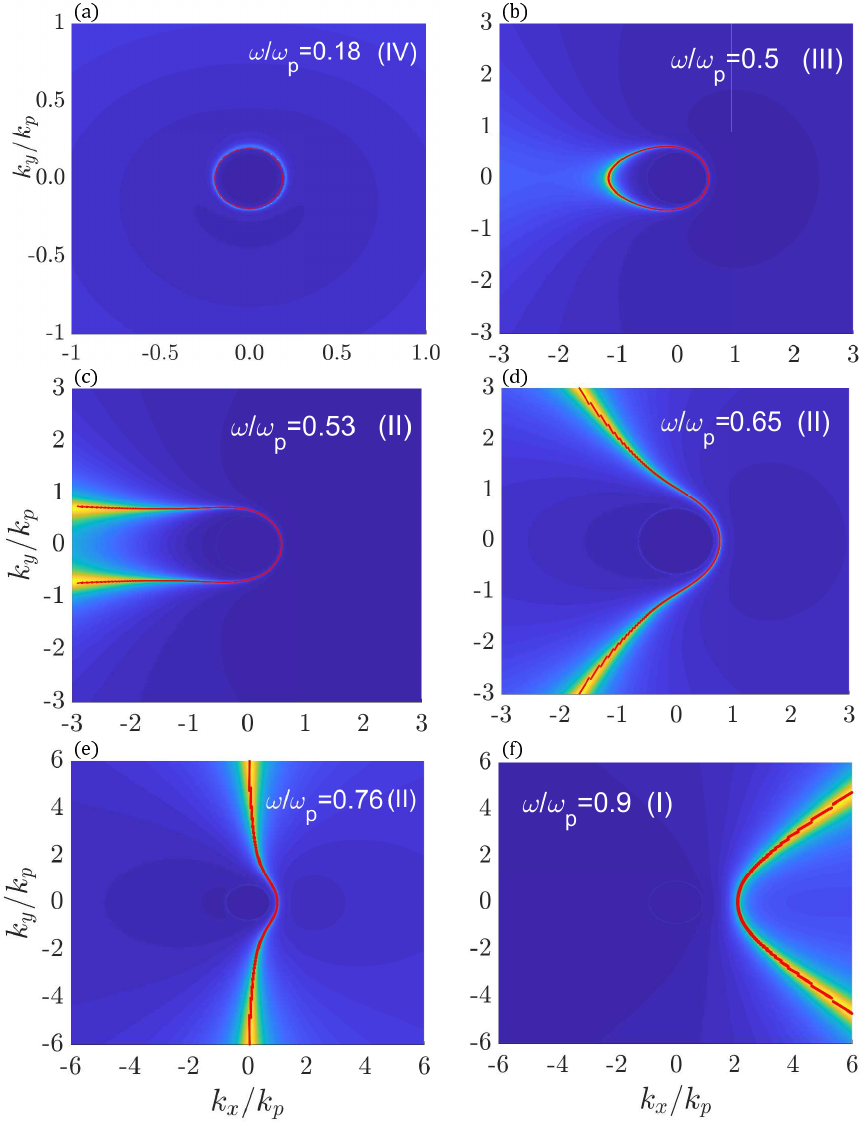}
	\caption{Density plot of the Sommerfeld integrand, $\left\vert F \right\vert $, (\ref{integrand}) and equi-frequency contours (solid red) extracted from (\protect\ref{SPPdisp}), for a biased-plasma-vacuum interface at different frequencies for $\Gamma /\protect\omega _{p}=0.015$. The notation I-IV refers to Fig. \ref{commonBG}.}
	\label{Density}
\end{figure}
Figure \ref{Density} shows several equi-frequency contours (EFC) of the dispersion surface at different frequencies (red lines). Also shown in Fig. \ref{Density} are density plots of the distribution function, $\left\vert F \right\vert $, obtained from the Green function and given by (\ref{integrand}). The phase and group velocities of an SPP are calculated as $\mathbf{v}_{p}=\mathbf{\hat{k}}_{s}\omega/\left\vert \mathbf{k}_{s}\right\vert$ and
\begin{equation}
	\mathbf{v}_{g}=\mathbf{\nabla}_{\mathbf{k}s}\omega\left(  \mathbf{k}%
	_{s}\right)  =\mathbf{\hat{x}}\frac{\partial\omega}{\partial k_{x}%
	}+\mathbf{\hat{y}}\frac{\partial\omega}{\partial k_{y}},
\end{equation}
respectively. This means that the group velocity, representing the directional flow of electromagnetic energy, is orthogonal to the equi-frequency contours. According to Fig. \ref{Density}a the EFCs at low frequencies are nearly circular such that energy flows isotropically. Hence, the resulting field pattern is essentially omni-directional (see Fig. \ref{polar}a discussed in the next section). As frequency increases, the semi-major axis of the EFC becomes elongated (Fig. \ref{Density}b) such that the energy begins to flow asymmetrically. For $\omega=0.53\omega_{p}$, the EFC becomes hyperbolic with the arms of the hyperbola widening as frequency increases (see Fig. \ref{Density}c-f). When the EFC becomes hyperbolic, two directional, narrow beams form in the SPP field pattern (see, e.g., Fig. \ref{polar}c,d). Moreover, the equi-frequency contours of the upper band in Fig. \ref{3DSPP}a show that the surface plasmons in this frequency range are mainly directed along the y direction (along the bias), existing down to the limit $k_{y}\rightarrow 0$.

\begin{figure}[!t]
	\includegraphics[width=0.99\columnwidth]{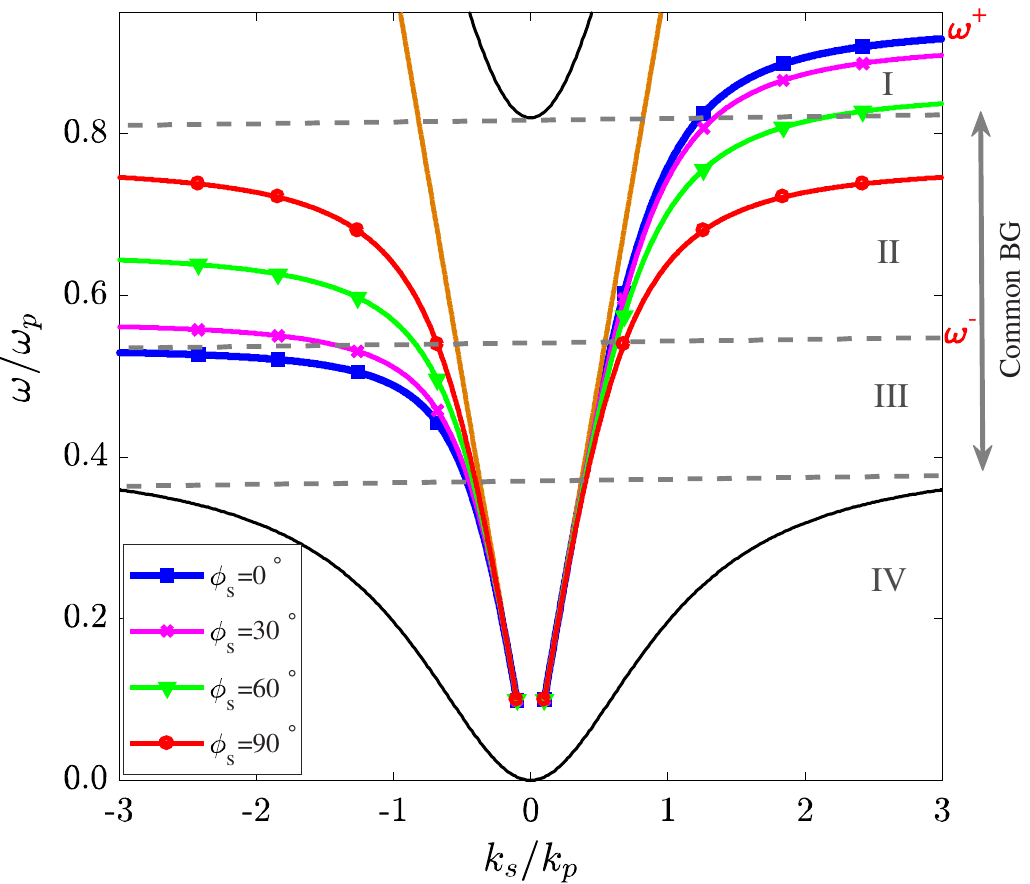}
	\caption{Two dimensional dispersion of the SPP for different propagation angles, $\protect\phi_{s}$, with respect to the positive (negative) x-axis for right (left) branches of the dispersion. The bulk dispersion (solid black) for $\alpha_{b}=0^{\circ}$ indicates the lower bandgap (BG), common to all propagation angles. The solid orange lines, symmetric with respect to $k_{s}=0$, show the dispersion of light in vacuum, e.g., $\omega/\omega_{p}= \pm k_{s}/k_{p}$.}
	\label{commonBG}
\end{figure}

Figure \ref{commonBG} shows the SPP dispersion behavior for the lower band, at different propagation angles (i.e., it shows several two dimensional traces of the SPP dispersion surface shown in Fig. \ref{3DSPP}b). Each branch of the SPP dispersion converges to \cite{3-PRA}
\begin{align}
	\omega_{\mathbf{k}}   &  =\frac{1}{2}\omega_{c}\cos
	\phi_{s}+\frac{1}{2}\sqrt{2\omega_{p}^{2}+\omega_{c}^{2}\left(  1+\sin^{2}\phi
	_{s}\right)  },
\end{align}
in the limit $k_{s}\rightarrow\infty$, derived using the quasi-static approximation. The maximum and minimum quasi-static resonance, $\omega^{\pm}=\omega_{k}\left(  \phi_{s}=0\right)  $, indicated in Figs. \ref{3DSPP}b and \ref{commonBG}, correspond to an SPP mode which propagates perpendicular to the bias. The dispersion is divided into four frequency regions: in Regions I and IV, there is no common bulk bandgap, whereas in Regions II and III, there exists a common bulk bandgap. In Region II, where the EFC is hyperbolic (see Fig. \ref{Density}c-f), we have directional propagation and the SPP field pattern consists of two narrow beams which are symmetric with respect to the $x$ axis (e.g. Fig. \ref{polar}c,d), and since $\omega(-k_{s})\neq\omega(k_{s})$, unidirectional behavior is also possible, making this frequency regime of central interest. Although in Region III there still exists a common bulk bandgap, narrow beams do not form in the SPP field pattern due to the fact that the EFC is ellipsoidal (see Fig. \ref{Density}b). Moreover, SPP propagation is nearly reciprocal. In Region IV, the EFC is circular (Fig. \ref{Density}a), indicating that the expected SPP field pattern is omni-directional (see Fig. \ref{polar}a), and from the dispersion shown in Fig. \ref{commonBG}, it is evident that the SPP is reciprocal, i.e. $\omega(-k_{s})=\omega(k_{s})$. 

As a partial summary, we have carefully studied the recently-identified lower band dispersion of surface waves on a dielectric-gyrotropic plasma interface, and have identified four regions (I-IV in Figs. \ref{3DSPP} and \ref{commonBG}) with different characteristics.

\section{Green function for a finite-thickness plasma, and SPP Beam Pattern in Space}
In the last section we considered a simple material-gyrotropic plasma interface, the Green function for which is provided in \cite{3-PRA}. In this section, we expand that analysis to consider a finite-thickness gyrotropic layer. We present a closed-form expression (as a Sommerfeld integral) for the Green function in the simple dielectric regions above and below the slab, which we believe to be a new result. Importantly, we also provide the Green function coefficient in quotient form for each case, which leads to the identification of the SPP dispersion equation (setting the denominator to zero), and allows the residue of the Green function, corresponding to the SPP, to be evaluated.

The procedure to derive the Green function follows that in \cite{3-PRA, 11-mario-optical}. The incident field excited by an electric dipole source, with dipole moment $\mathbf{p}_{e}=\mathbf{\hat{x}}p_{x}+\mathbf{\hat{y}}p_{y}+\mathbf{\hat{z}}p_{z}$, suspended a distance $d$ above the first interface, is given by $\mathbf{E}^{p}\left(  \mathbf{r}\right)  =\left(  \mathbf{\nabla\nabla}+\mathbf{\bar{I}}k_{0}^{2}\varepsilon_{r,0}\right)  \cdot\mathbf{\pi}^{p}\left(  \mathbf{r}\right)$, where $\mathbf{\pi}^{p}\left(  \mathbf{r}\right)  $ denotes the principal hertzian potential due to the dipole source, which we write in terms of the principal Green function, $\mathbf{\pi}^{p}\left(  \mathbf{r}\right)  =g^{p}\left(\mathbf{r,r}_{0}\right)  \mathbf{p}_{e}/\varepsilon_{0}\varepsilon_{r,0}$, where $g^{p}\left(\mathbf{r,r}_{0}\right)  =e^{j k_{0}\sqrt{\varepsilon_{r,0}}\left\vert \mathbf{r}-\mathbf{r}_{0}\right\vert}/4\pi\left\vert \mathbf{r}-\mathbf{r}_{0}\right\vert $ such that $\varepsilon_{r,0}$ is the relative permitivitty of the top layer (see Fig. \ref{geom}) and $\mathbf{r}_{0}=(0,0,d)$. Following \cite{3-PRA}, the principal and scattered fields may be written similarly in Sommerfeld integral form,
\begin{align}
	\mathbf{E}^{p}\left(  \mathbf{r}\right)   &  =\int d^{2}\mathbf{k}%
	_{s}e^{j\mathbf{k}_{s}\cdot\mathbf{r}}\frac{e^{-\gamma\left\vert
	z-z_{0}\right\vert }}{8\pi^{2}\varepsilon_{0}\varepsilon_{r,0}\gamma
	}\mathbf{\bar{C}}_{z \gtrless d}^{p}\cdot\mathbf{p,}\label{r14}\\
	\mathbf{E}^{r}\left(  \mathbf{r}\right)   &  =\int d^{2}\mathbf{k}%
	_{s}e^{j\mathbf{k}_{s}\cdot\mathbf{r}}\frac{e^{-\gamma\left(  z+z_{0}\right)
	}}{8\pi^{2}\varepsilon_{0}\varepsilon_{r,0}\gamma}\mathbf{\bar{C}}^{r}%
	\cdot\mathbf{p,}\label{r15}\\
	\mathbf{E}^{t}\left(  \mathbf{r}\right)   &  =\int d^{2}\mathbf{k}%
	_{s}e^{j\mathbf{k}_{s}\cdot\mathbf{r}}\frac{e^{\gamma\left(  z-z_{0}\right)
	}}{8\pi^{2}\varepsilon_{0}\varepsilon_{r,0}\gamma}\mathbf{\bar{C}}^{t}%
	\cdot\mathbf{p,} \label{r16}%
\end{align}
where $\mathbf{\bar{C}}_{z \gtrless d}^{p}$ and $\mathbf{\bar{C}}^{r,t}$ take the form,
\begin{align}
	\mathbf{\bar{C}}_{z \gtrless d}^{p}  &  =\mathbf{\bar{A}}_{z \gtrless d}%
	\cdot\mathbf{\bar{I}}_{s}\cdot\mathbf{\bar{B},}\label{r17}\\
	\mathbf{\bar{C}}^{r,t}  &  =\mathbf{\bar{A}}^{r,t}\cdot\left\{
	\mathbf{\bar{R},\bar{T}}\right\}  \cdot\mathbf{\bar{B},} \label{r18}%
\end{align}
such that
\begin{align}
	\mathbf{\bar{A}}_{z \gtrless d}  &  =\mathbf{\bar{I}}_{s}\mp\frac{1}{k_{z}%
	}\mathbf{\hat{z}k}_{s},\label{r19}\\
	\mathbf{\bar{A}}^{r,t}  &  =\mathbf{\bar{I}}_{s}\mp\frac{1}{k_{z}^{r,t}%
	}\mathbf{\hat{z}k}_{s},\label{r20}\\
	\mathbf{\bar{B}}  &  = k_{0}^{2}\varepsilon_{r,0}\mathbf{\bar{I}}_{s}-\mathbf{k}_{s}%
	\mathbf{k}_{s}+k_{z}\mathbf{k}_{s}\mathbf{\hat{z},} \label{r21}%
\end{align}
where $\mathbf{\bar{I}}_{s}=\mathbf{\hat{x}\hat{x}}+\mathbf{\hat{y}\hat{y}}$, $k_{z}=k_{z}^{r}=\sqrt{k_{0}^{2}\varepsilon_{r,0}-k_{x}^{2}-k_{y}^{2}}$, and $k_{z}^{t}=\sqrt{k_{0}^{2}\varepsilon_{r,2}-k_{x}^{2}-k_{y}^{2}}$. The reflection and transmission coefficients for a slab of finite depth, $h$, are denoted by $\mathbf{\bar{R}}\left(  \omega,\mathbf{k}_{s}\right)  $ and $\mathbf{\bar{T}}\left(\omega,\mathbf{k}_{s}\right)  $, respectively. It is shown in the appendix that these $2\times2$ tensor coefficients take the form
\begin{align}
	\mathbf{\bar{R}}  &  =\mathbf{\bar{R}}_{01}+\mathbf{\bar{T}}_{10}%
	\cdot\mathbf{\bar{R}}_{12}^{\prime}\cdot\left(  \mathbf{\bar{I}}_{s}%
	-\mathbf{\bar{R}}_{10}\cdot\mathbf{\bar{R}}_{12}^{\prime}\right)  ^{-1}%
	\cdot\mathbf{\bar{T}}_{01},\label{r22}\\
	\mathbf{\bar{T}}  &  =\mathbf{\bar{T}}_{12}\cdot\mathbf{\bar{P}}_{E}^{-}%
	\cdot\left(  \mathbf{\bar{I}}_{s}-\mathbf{\bar{R}}_{10}\cdot\mathbf{\bar{R}}%
	_{12}^{\prime}\right)  ^{-1}\cdot\mathbf{\bar{T}}_{01}, \label{r23}%
\end{align}
where $\mathbf{\bar{T}}_{nn^{\prime }}=\mathbf{\bar{I}}_{s}+\mathbf{\bar{R}}_{nn^{\prime }}$ for $\left(  n,n^{\prime }\right)  \in\left\{  \left(  0,1\right)  ,\left(1,0\right)  ,\left(  1,2\right)  \right\} $ and
\begin{equation}
	\mathbf{\bar{R}}_{12}^{\prime}=\mathbf{\bar{P}}_{E}^{+}\cdot\mathbf{\bar{R}%
	}_{12}\cdot\mathbf{\bar{P}}_{E}^{-},\label{r24}%
\end{equation}
such that $\mathbf{\bar{P}}_{E}^{m}$ denotes the spacial propagator, which accounts for the accumulated phase as the wave propagates within the gyrotropic medium in the $\pm z$ directions. The single interface reflection coefficients associated with each interface, $\mathbf{\bar{R}}_{nn^{\prime }}$, along with the spacial propagator, $\mathbf{\bar{P}}_{E}^{m}$, can alternatively be expressed in numerator/denominator form as%
\begin{align}
	\mathbf{\bar{R}}_{nn^{\prime }} &  =\frac{1}{k_{y}\Omega^{nn^{\prime }}}\left(
	\begin{array}
	[c]{cc}%
	k_{y}\Pi_{11}^{nn^{\prime }} & \Pi_{12}^{nn^{\prime }}\\
	k_{y}^{2}\Pi_{21}^{nn^{\prime }} & k_{y}\Pi_{22}^{nn^{\prime }}%
	\end{array}
	\right)  ,\label{r25}\\
	\mathbf{\bar{P}}_{E}^{m} &  =\frac{1}{k_{y}\chi^{m}}\left(
	\begin{array}
	[c]{cc}%
	k_{y}\Delta_{11}^{m} & \Delta_{12}^{m}\\
	k_{y}^{2}\Delta_{21} & k_{y}\Delta_{22}^{m}%
	\end{array}
	\right)  ,\label{r26}
\end{align}
where the quantities $\Omega^{nn^{\prime }}$, $\Pi^{nn^{\prime }}$, $\chi^{m}$, and $\Delta^{m}$ are defined in the appendix. For the single interface case, we find that setting $\Omega^{01}$ to zero in $\mathbf{\bar{R}}_{01}$ gives the expected dispersion relation for the SPP (\ref{SPPdisp}).
\begin{figure}[!t]
	\includegraphics[width=0.99\columnwidth]{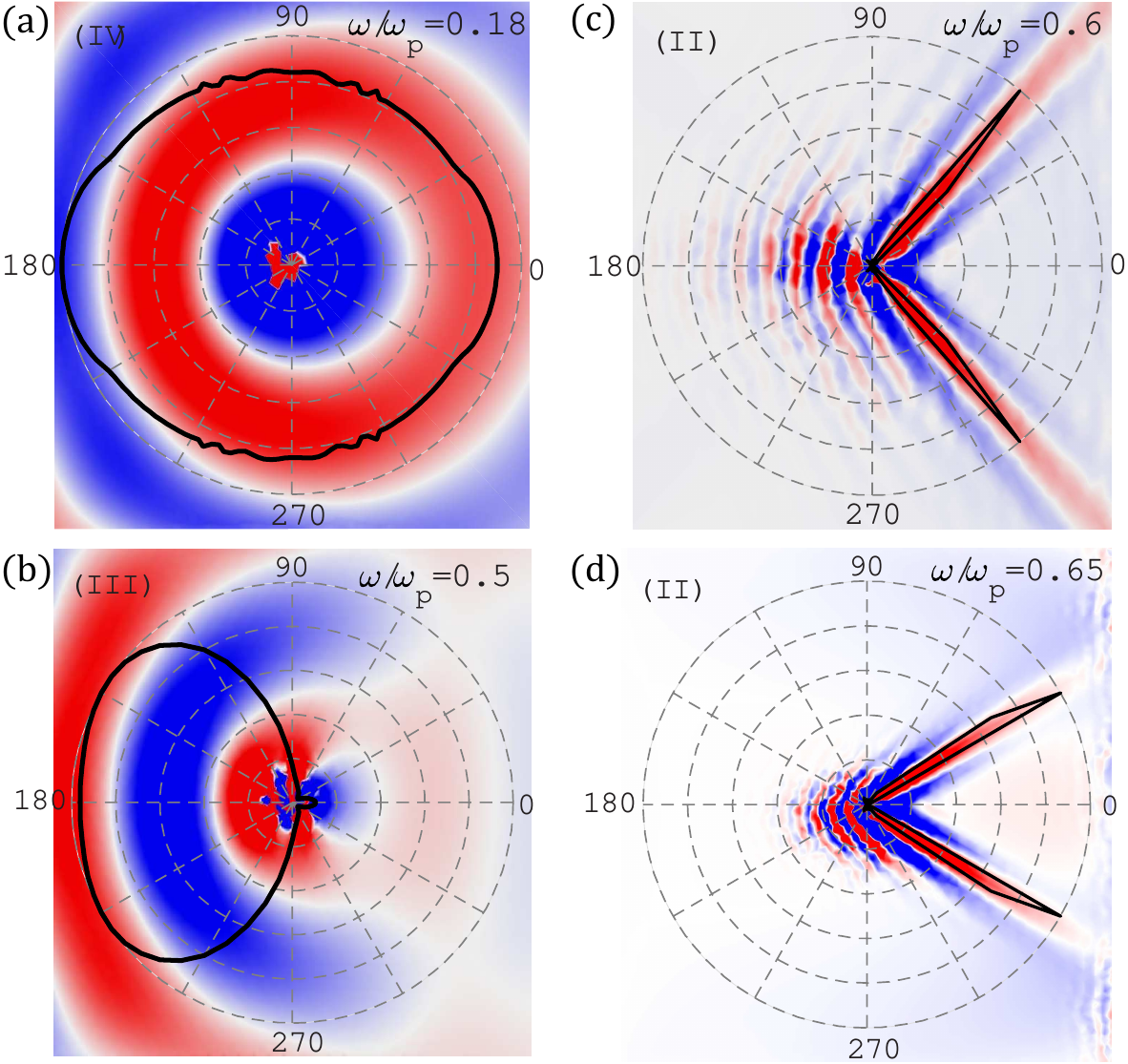}
	\caption{Scattered electric field, $\left| E_{z}^{r} \right|$, obtained from the Green function (solid black lines) for a biased-plasma-vacuum interface, with $\rho=0.7\protect\lambda$,  $z=0.008\protect\lambda _{p}$, where $\protect\lambda=2\protect\pi c/\omega$ and $\protect\lambda _{p}=2\protect\pi c/\protect\omega _{p}$. For comparison, the electric field distribution generated using COMSOL is also shown.}
	\label{polar}
\end{figure}
\begin{figure*}[tbh]
	\includegraphics[width=1.99\columnwidth]{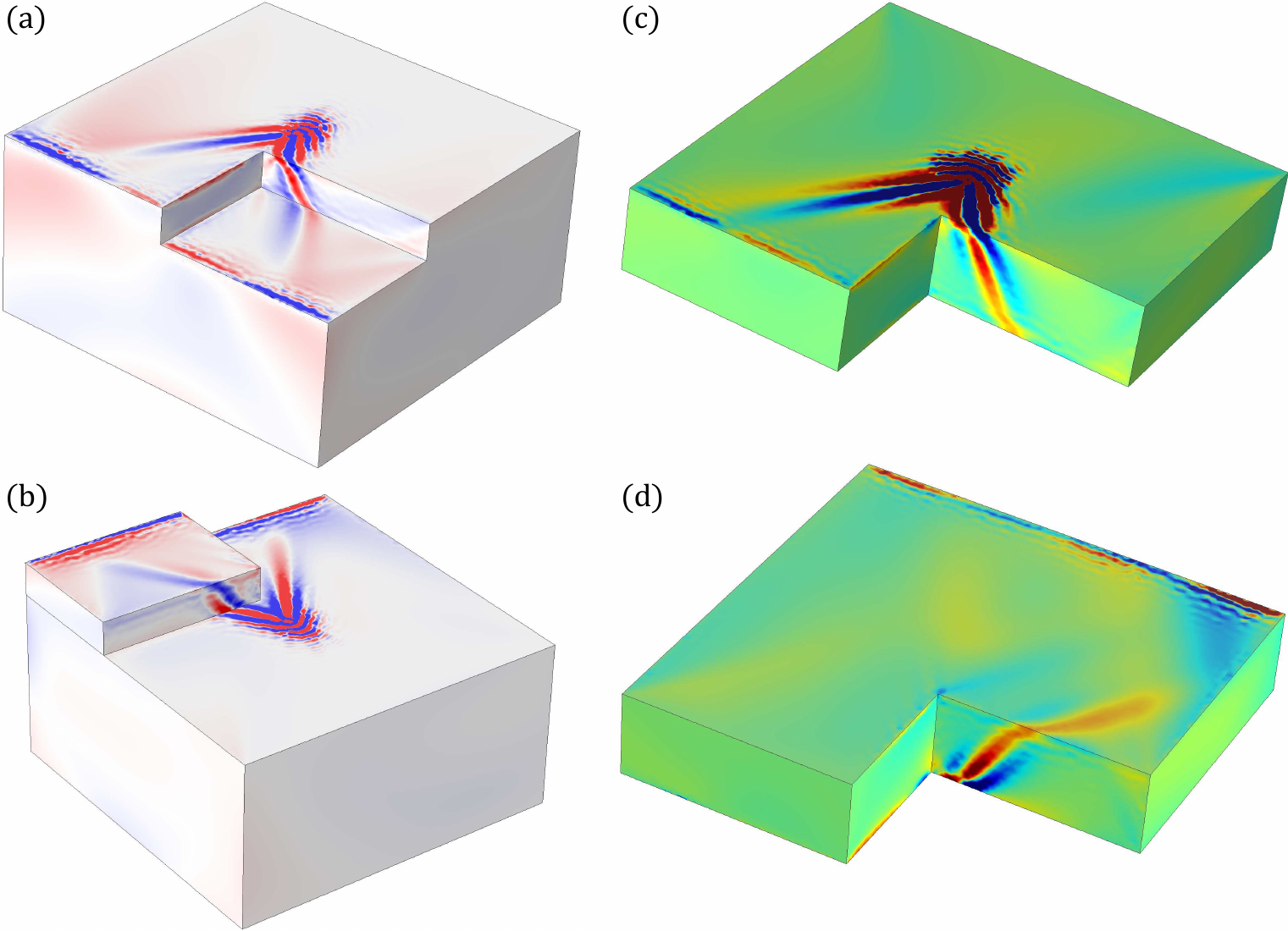}
	\caption{(a,b) Electric field (computed using COMSOL) at the interface of a thick (essentially infinite) gyrotropic plasma slab in the presence of (a) a hole discontinuity and (b) a block discontinuity. (c,d) Electric field at the top (c) and bottom (d) of a finite thickness slab ($h=0.12\lambda_{p}$) in the presence of a discontinuity. The SPP, excited on the top interface by a point dipole, propagates around the open surface to the bottom side of the plasma.}
	\label{discon1}
\end{figure*}

In the special case where a $z$ directed dipole moment, $\mathbf{p}_{e}=\mathbf{\hat{z}}p_{z}$, is placed at a height $d$ above the first interface ($z>0$), the $z$ component of the scattered electric field simplifies to
\begin{equation}
E_{z}^{r}\left( \mathbf{r}\right) =\int d^{2}\mathbf{k}_{s}F(\mathbf{k}_{s},%
\mathbf{r},\omega ),  \label{GreenFn}
\end{equation}
where
\begin{equation}
	F\left(\mathbf{k}_{s},\mathbf{r},\omega \right)=e^{j\mathbf{k}_{s}\cdot\mathbf{r}}\frac{e^{-\gamma\left(  z+d\right)}}{8\pi^{2}\varepsilon_{0}\varepsilon_{r,0}\gamma}		C_{zz}^{r}p_{z},\label{integrand}
\end{equation}
such that, for a single interface,%
\begin{equation}
	C_{zz}^{r}=\frac{-k_{x}\left(  \Pi_{12}^{01}+k_{x}\Pi_{11}^{01}\right)
	}{\Omega^{01}}-\frac{k_{y}^{2}\left(  \Pi_{22}^{01}+k_{x}\Pi_{21}^{01}\right)
	}{\Omega^{01}}.%
	\label{Czz}
\end{equation}

Using (\ref{GreenFn}), the electric field distribution near the interface of a half-space gyrotropic media for $\rho=0.7\lambda$, $z=0.008\lambda_{p}$, and $0<\phi<2\pi$, is shown in Fig. \ref{polar}. The results obtained in COMSOL are also shown Fig. \ref{polar}, and agree with the Green function analysis.

As shown in Fig. \ref{polar}a, the expected behavior of surface wave propagation for operating frequencies that lie in Region IV of the dispersion (see Figs. 3 and 5), is omnidirectional. In Region III, propagation is bi-directional, with the SPP intensity concentrated to one half plane as depicted in Fig. \ref{polar}b. Transitioning from Region IV to Region I, the expected behavior increasingly tends toward unidirectional. Interestingly, for frequencies that satisfy the SPP resonant condition, $\omega^{-}<\omega<\omega^{+}$ (Regions I and II), Fig. \ref{polar}c,d, show that narrow-beam directional propagation is obtained, consistent with the previous discussion of equi-frequency contours; two representative results which satisfy the resonant condition, $\omega=0.6\omega_{p}$ and $\omega=0.65\omega_{p}$ are shown. At $\omega=\omega^{-}$, the field pattern forms two narrow beams which approach each other as the operating frequency increases. Eventually, the two beams join to form a single beam at $\omega=0.76\omega_{p}$, corresponding to the saturation frequency of the $\phi_{s}=90^{\circ}$ branch in Fig. \ref{commonBG}, and then split to form two beams for $0.76\omega_{p}<\omega<\omega^{+}$. Therefore, the angle of the beams with respect to the $x$ axis is adjustable with frequency as well as the magnetic bias. Furthermore, if the direction of the magnetic bias is flipped, the beams propagate in the opposite direction.

To have an indication of the inherent robustness of the SPP within the resonant range, a discontinuity in the form of a hole/block is constructed in an attempt to impede the SPP. A unidirectional SPP that crosses a band gap in reciprocal space is immune to the effects of back-scattering and diffraction. To illustrate this, Fig. \ref{discon1}a,b shows the electric field due to a electric point source near the vacuum-plasma interface of a plasma half-space. The SPP passes through the discontinuity without reflection or diffraction. Similarly, for a finite-thickness slab, the SPP excited on the top surface, upon encountering the end of the plasma, passes onto the bottom surface, as shown in Fig.  \ref{discon1}c (top view) and Fig.  \ref{discon1}d (bottom view).

\begin{figure}[!bth]
	\includegraphics[width=0.99\columnwidth]{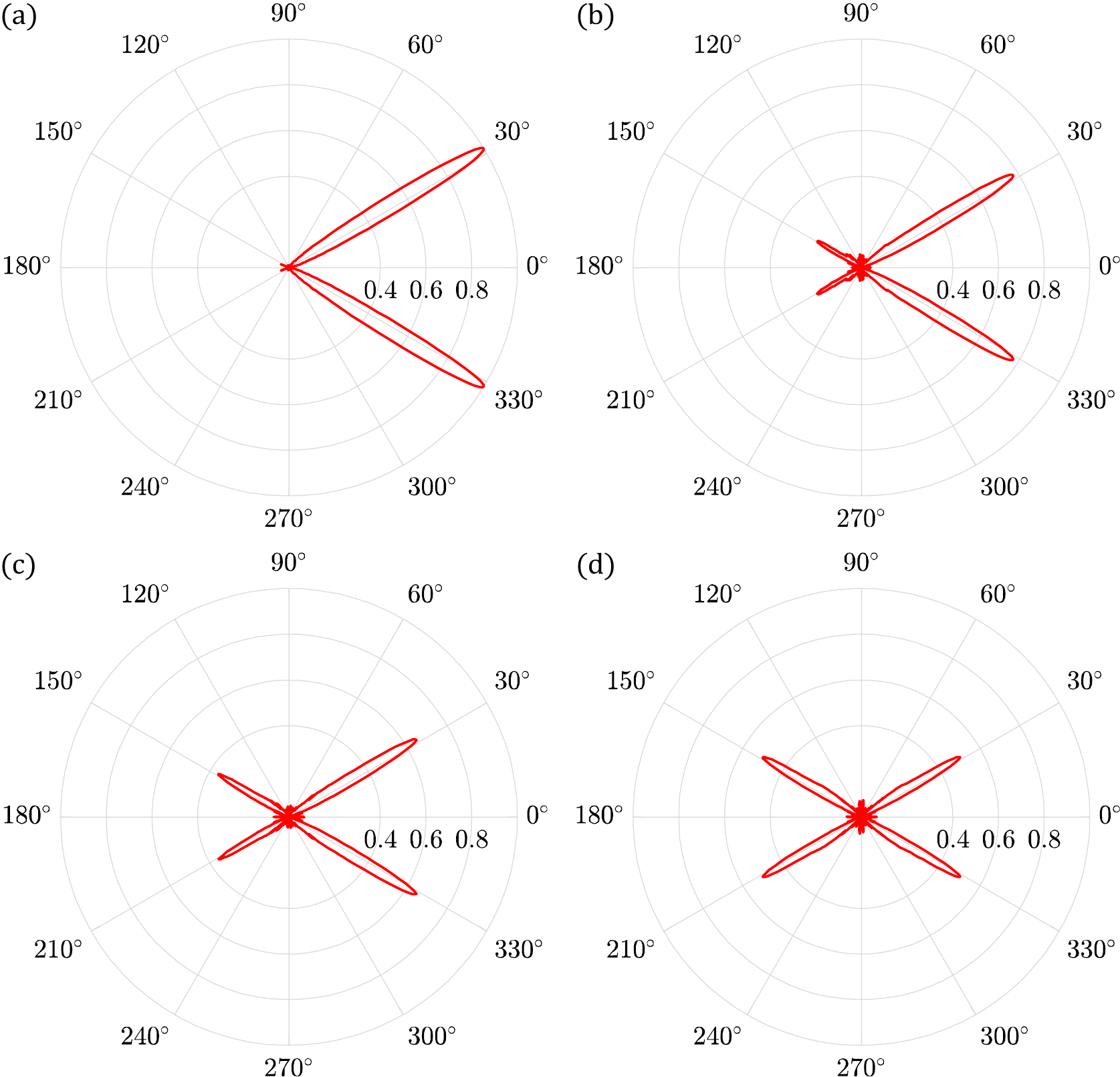}
	\caption{SPP beam pattern excited by a vertical dipole source at the interface of a finite slab of thickness $h = \lambda_{p}$, obtained by evaluating (\ref{r15}) for set observation height, $z = 0.008\lambda_{p}$ and in-plane radial distance, $\rho=0.08\lambda_{p}$, $\lambda_{p} = 2 \pi c / \omega_{p}$. Four values of loss are considered such that $\Gamma= 10^{-4}\omega_{p}$ (a), $\Gamma=10^{-5}\omega_{p}$ (b), $\Gamma=30^{-6}\omega_{p}$ (c), and $\Gamma=0$ (d). These results are normalized with respect to the beam maximum extracted from the field profile shown in (a).}\label{sppvsloss}
\end{figure}

As shown above, the vacuum-plasma interface can support a uni-directional SPP. However, it is not clear if a thin, finite-thickness slab can also support such an SPP. Figure \ref{sppvsloss} shows the SPP pattern obtained by evaluating the scattered/reflected Green function field (\ref{r15}) as a function of angular position in the xoy plane. For this analysis, we consider a vertical dipole source, operating with frequency $\omega=0.65\omega_{p}$ and positioned at the upper interface ($z_{1}=0$) of a gyrotropic plasma slab with a fixed thickness $h=\lambda_{p}$. Figure \ref{sppvsloss} shows the scattered field for the fixed observation point $\left(\rho,z\right)=\left(0.08\lambda_{p},0.008\lambda_{p}\right)$ and several values of loss within the range $0<\Gamma<10^{-4}\omega_{p}$. For a sufficient amount of loss, $\Gamma=10^{-4}\omega_{p}$, only two beams appear in the field pattern, similar to those obtained for a single interface (see Fig. \ref{polar}d). As the loss decreases from $\Gamma= 10^{-4}\omega_{p}$ to $\Gamma= 0$, we see the emergence of two backward beams present on the upper interface (due to the evanescent tail of the bottom-surface SPP), which indicates the breakdown of uni-directional behavior. 
\subsection{Quasi-Static Approximation}\label{quasi1}
Further insight can be gained by a quasi static approximation, where the electric field is written in terms of the electro-static potential, $\phi_{k}$, such that $E_{k}\approx-\nabla\phi_{k}$, assuming the associated magnetic field is negligible. Solving Gauss' law in both isotropic and gyrotropic media, and applying boundary conditions for the tangential components of the electric field at each interface, the electric potential for a symmetric slab (centered at $z=0$) is obtained as
\begin{widetext}
	\begin{equation}
		\phi_{k}=e^{j\mathbf{k}_{s}\cdot\mathbf{r}}\left\{
		\begin{array}
		[c]{cc}%
		\left[  jC_{1}\sinh\left(  \tilde{k}_{s}h/2\right)  +C_{2}\cosh\left(
		\tilde{k}_{s}h/2\right)  \right]  e^{-k_{s}\left(  z-h/2\right)  } & z>h/2\\
		jC_{1}\sinh\left(  \tilde{k}_{s}z\right)  +C_{2}\cosh\left(  \tilde{k}%
		_{s}z\right)  & -h/2<z<h/2\\
		\left[  -jC_{1}\sinh\left(  \tilde{k}_{s}h/2\right)  +C_{2}\cosh\left(
		\tilde{k}_{s}h/2\right)  \right]  e^{-k_{s}\left(  z-h/2\right)  } & z<-h/2
		\end{array}
		\right., \label{r29}%
	\end{equation}
\end{widetext}
where $\tilde{k}_{s}=\sqrt{k_{x}^{2}+\varepsilon_{a}k_{y}^{2}/\varepsilon_{t}}$, $h$ denotes the slab thickness and $C_{1}$ and $C_{2}$ are parameters can be obtained by applying the mode orthogonality condition. Enforcing continuity of the normal components of electric displacement at the two interfaces leads to the quasi-static SPP dispersion relation
\begin{equation}
	\varepsilon_{g}^{2}k_{x}^{2}-\varepsilon_{t}^{2}\tilde{k}_{s}^{2}%
	-k_{s}^{2}=2\varepsilon_{t}k_{s}%
	\tilde{k}_{s}\coth\left(  \tilde{k}_{s}h\right). \label{quasidisp}
\end{equation}
The quasi-static approximation is valid only for SPPs with short wavelength ($k_{s}\rightarrow\infty$). In the limit $h\rightarrow\infty$, \ the dispersion relation reduces to that derived for a single interface \cite{3-PRA},
\begin{equation}
	k_{s}+k_{x}\varepsilon_{g}+\tilde{k}_{s}\varepsilon_{t}=0. \label{quasidisp2}
\end{equation} 

\begin{figure}[!htb]\includegraphics[width=0.99\columnwidth]{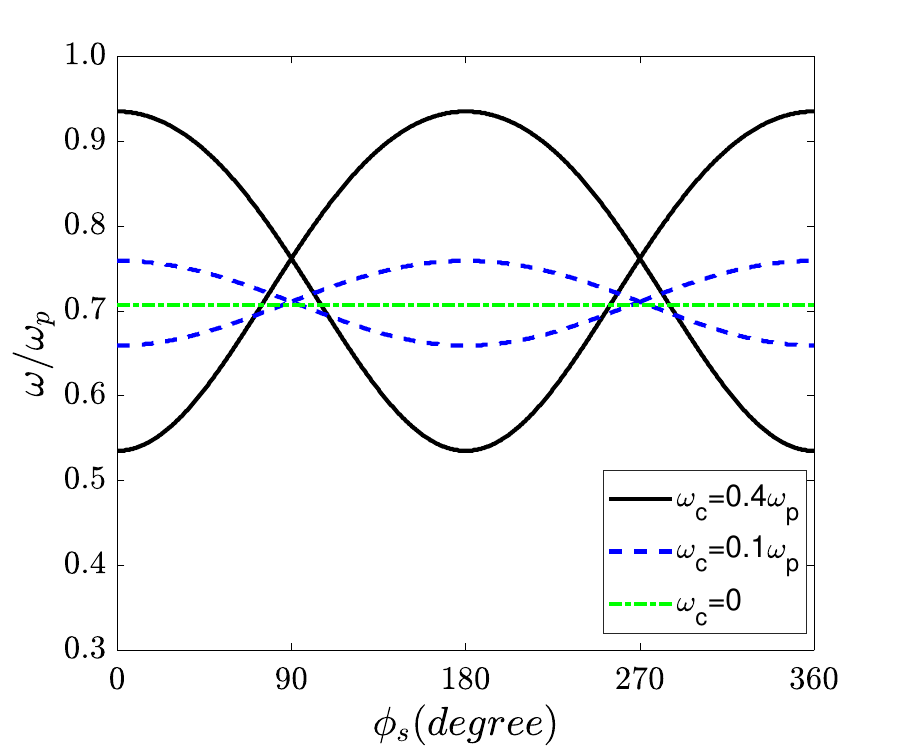}
	\caption{Solutions to the quasi-static SPP dispersion relation (\ref{quasidisp}) for a finite thickness slab of thickness $h=0.25\lambda_p$ and wavenumber $k_{s}= 10k_{p} \gg 1/h$. The cyclotron frequency ranges from $0$ to $0.4\omega_{p}$. From these results, we find that for a given operation frequency, a maximum of four beams is possible in the SPP beam pattern. Additionally, we find that as magnetic bias increases, the SPP resonant range also increases.}
	\label{Quasi}
\end{figure}

Figure \ref{Quasi} shows the solutions to the quasi-static relation (\ref{quasidisp}) for several values of cyclotron frequency, representing the SPP resonance in the quasi-static limit. For a given $\omega$ value, there are four values of $\phi_{s}$, two of which correspond to the forward beams and the other two correspond to the backward beams (see Fig. \ref{sppvsloss}). In the presence of a magnetic bias, the SPP resonance depends on the direction of the SPP modes, however, it is independent of the slab thickness for large values of $k_{s}$. Numerically we find that in the absence of magnetic bias ($\omega_{c}=0$), the SPP resonance at $\underset{\omega _{c}\rightarrow 0}{\lim }\omega _{SPP}=\omega _{p}/\sqrt{2}$, which shows that SPPs become direction independent in this limit, as expected.

The quasi-static dispersion in Fig. \ref{Quasi} suggests that four beams may be present in the scattered field profile for operating frequencies that fall within the SPP resonant range $\omega^{-}<\omega<\omega^{+}$. For example, consider an operating frequency of $\omega = 0.65\omega_{p}$ and cyclotron frequency $\omega_{c} = 0.4\omega_{p}$. From the quasi static dispersion, we find that the in-plane wave vector, and hence, phase velocity, of the SPP (approximately) makes an angle $\phi_{s} \in \left\{ 60^{\circ},120^{\circ},240^{\circ},300^{\circ} \right\}$ with respect to the x-axis. The group velocity (i.e. the direction of energy flow as indicated by the direction of the beams) of the SPP is perpendicular to the phase velocity and therefore, makes an angle $\phi_{s}+90^{\circ} \in \left\{ 150^{\circ},210^{\circ},330^{\circ},30^{\circ}\right\}$ with respect to the x-axis. In the low loss limit, the scattered field profile shows four beams with the expected aforementioned angles made with respect to the x-axis (see Fig. \ref{sppvsloss}d). However, for a lossy slab, we find that only two beams become present on any given surface at angles $\phi_{s}+90^{\circ} \in \left\{ 330^{\circ},30^{\circ}\right\}$ (top) and $\phi_{s}+90^{\circ} \in \left\{ 150^{\circ},210^{\circ}\right\}$ (bottom) (see Fig. \ref{sppvsloss}a). That is, the quasi-static analysis provides four symmetric beams, two of which will be excited on a given interface (top or bottom).

\section{Conclusion}
We have investigated the behavior of surface plasmon polaritons propagating at the interface between vacuum and gyrotropic plasma for both infinite- and finite-thickness slab configurations. We have identified a bulk bandgap, common to all propagation angles. The operating frequency is chosen to lie within the lower common band gap, wherein omni-directional, bidirectional, and narrow directional beam patterns are observed. Operating in the bandgap gives the SPP interesting properties that protect it from back scatter and diffraction in the presence of a discontinuity. The direction of the SPP beams are adjustable with operation frequency and also the bias magnetic field. The Green function and quasi-static approximation to the dispersion have also been obtained for a finite-thickness slab.\label{SectConcl}
\section*{Appendix: Dyadic Green function for a finite thickness slab}\label{ApGreenHalfSpace}
\begin{figure}[!thb]
	\includegraphics[width=0.95\columnwidth]{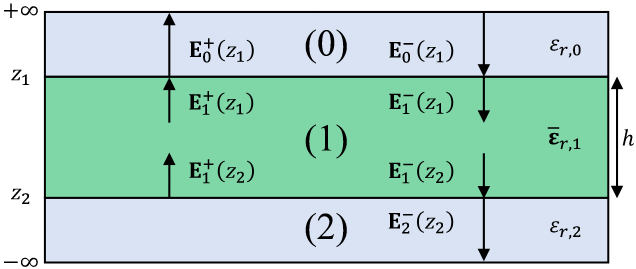}
	\caption{Cross sectional view of Fig. \ref{geom}. The top and bottom interfaces are positioned at $z=z_{1} = 0$ and $z=z_{2} = -h$ respectively. Regions (0) and (2) are characterized by $\varepsilon_{r,0}$ and $\varepsilon_{r,2}$ respectively, while Region (1) is characterized by the gyrotropic permitivitty tensor, $\bar{\varepsilon}_{r,1}$, defined in (\ref{r1}). The electric fields associated with plane waves propagating in each region, with group velocity in the $\pm z$ directions, are also shown.}
\end{figure}
Here, we derive the plane wave reflection and transmission coefficients which relate the tangential field components of the electric field reflected and transmitted from a gyrotropic slab of finite thickness, $h$. As in \cite{3-PRA}, it is important to define a convenient, orthogonal coordinate system in which to expand the amplitude vector of a plane wave propagating in the gyrotropic medium. The set of orthogonal unit vectors which span this coordinate system is given by $\left\{  \mathbf{\hat{k}}_{t,i}^{m},\mathbf{\hat{y}},\mathbf{\hat{k}}_{t,i}^{m}\times\mathbf{\hat{y}}\right\}  $, where $\mathbf{\hat{k}}_{t,i}^{m}=\mathbf{\hat{x}}k_{x}+\mathbf{\hat{z}}mk_{z,i}$ for $m\in\left\{  \pm\right\}$ and $i\in\left\{1,2\right\}$. The fields above and below the interface, are simply expanded in terms of the Cartesian basis, $\left\{  \mathbf{\hat{x}},\mathbf{\hat{y}},\mathbf{\hat{z}}\right\}$. The relationship between the electric and magnetic fields above and below the slab is given by
\begin{equation}
	\left(
	\begin{array}
	[c]{c}%
	\omega\mu_{0}H_{y}^{m}\\
	\omega\mu_{0}H_{x}^{m}%
	\end{array}
	\right)  =\left\{  \mathbf{\bar{Y}}^{m},\mathbf{\bar{Y}}_{g}		^{m}\right\}
	\cdot\left(
	\begin{array}
	[c]{c}%
	E_{x}^{m}\\
	E_{y}^{m}%
	\end{array}
	\right)  , \label{r40}%
\end{equation}
where the electric and magnetic fields in the dielectric regions are related using%
\begin{equation}
	\mathbf{\bar{Y}}^{m}=\frac{1}{mk_{z}}\left(
	\begin{array}
	[c]{cc}%
	\left(  k_{x}^{2}+k_{z}^{2}\right)   & k_{x}k_{y}\\
	-k_{x}k_{y} & -\left(  k_{y}^{2}+k_{z}^{2}\right)
	\end{array}
	\right)  ,\label{r50}%
\end{equation}
while the electric and magnetic fields within the gyrotropic plasma are related using
\begin{align}
	\mathbf{Y}_{g}^{m}=\left(
	\begin{array}
	[c]{cc}%
	-\delta_{1}k_{t,1}^{2} & -\delta_{2}k_{t,2}^{2}\\
	k_{y}\phi_{1}^{m} & k_{y}\phi_{2}^{m}%
	\end{array}
	\right)  \cdot\left(
	\begin{array}
	[c]{cc}%
	\beta_{1}^{m} & \beta_{2}^{m}\\
	k_{y}\theta_{1} & k_{y}\theta_{2}%
	\end{array}
	\right)  ^{-1}.\label{rr}%
\end{align}
Matching the tangential components of the electric and magnetic fields at each interface yields
\begin{align}
	\mathbf{\bar{T}}_{01}\cdot\mathbf{E}_{0}^{-}\left(  z_{1}\right)   &  =\left(
	\mathbf{\bar{I}}_{s}+\mathbf{\bar{R}}_{01}\right)  \cdot\mathbf{E}_{0}%
	^{-}\left(  z_{1}\right),\label{r61}\\
	\mathbf{\bar{T}}_{10}\cdot\mathbf{E}_{1}^{+}\left(  z_{1}\right)   &  =\left(
	\mathbf{\bar{I}}_{s}+\mathbf{\bar{R}}_{10}\right)  \cdot\mathbf{E}_{1}%
	^{+}\left(  z_{1}\right),\label{r62}\\
	\mathbf{\bar{T}}_{12}\cdot\mathbf{E}_{1}^{-}\left(  z_{2}\right)   &  =\left(
	\mathbf{\bar{I}}_{s}+\mathbf{\bar{R}}_{12}\right)  \cdot\mathbf{E}_{1}%
	^{-}\left(  z_{2}\right),\label{r63}\\
	\mathbf{Y}_{g}^{-}\cdot\mathbf{\bar{T}}_{01}\cdot\mathbf{E}_{0}^{-}\left(
	z_{1}\right)   &  =\mathbf{\bar{Y}}^{-}\cdot\mathbf{E}_{0}^{-}\left(
	z_{1}\right)  \nonumber\\
	&  +\mathbf{\bar{Y}}^{+}\cdot\mathbf{\bar{R}}_{01}\cdot\mathbf{E}_{0}%
	^{-}\left(  z_{1}\right),\label{r64}\\
	\mathbf{Y}^{+}\cdot\mathbf{\bar{T}}_{10}\cdot\mathbf{E}_{1}^{+}\left(
	z_{1}\right)   &  =\mathbf{\bar{Y}}_{g}^{+}\cdot\mathbf{E}_{1}^{+}\left(
	z_{1}\right)  \nonumber\\
	&  +\mathbf{\bar{Y}}_{g}^{-}\cdot\mathbf{\bar{R}}_{10}\cdot\mathbf{E}_{1}%
	^{+}\left(  z_{1}\right),\label{r65}\\
	\mathbf{Y}^{-}\cdot\mathbf{\bar{T}}_{12}\cdot\mathbf{E}_{1}^{-}\left(
	z_{2}\right)   &  =\mathbf{\bar{Y}}_{g}^{-}\cdot\mathbf{E}_{1}^{-}\left(
	z_{2}\right)  \nonumber\\
	&  +\mathbf{\bar{Y}}_{g}^{+}\cdot\mathbf{\bar{R}}_{12}\cdot\mathbf{E}_{1}%
	^{-}\left(  z_{2}\right).\label{r66}%
\end{align}
From (\ref{r61})-(\ref{r66}) we find $\mathbf{\bar{T}}_{nn^{\prime }}=\mathbf{\bar{I}%
}_{s}+\mathbf{\bar{R}}_{nn^{\prime }}$ where%
\begin{equation}
	\mathbf{\bar{R}}_{nn^{\prime}}=\left(  \mathbf{\bar{Y}}^{m_{1}}-\mathbf{\bar{Y}}%
	_{g}^{m_{2}}\right)  ^{-1}\cdot\left(  \mathbf{\bar{Y}}_{g}^{m_{3}%
	}-\mathbf{\bar{Y}}^{m_{3}}\right)  , \label{r67}%
\end{equation}
such that
\begin{equation}
	\left(  m_{1},m_{2},m_{3}\right)  =\left\{
	\begin{array}
	[c]{cc}%
	\left(  +,-,-\right)   & \left(  n,n^{\prime}\right)  =\left(  0,1\right)  \\
	\left(  +,-,+\right)   & \left(  n,n^{\prime}\right)  =\left(  1,0\right)  \\
	\left(  -,+,-\right)   & \left(  n,n^{\prime}\right)  =\left(  1,2\right)
	\end{array}
	\right.  .\label{r69}%
\end{equation}
Furthermore, it is noted that%
\begin{align}
	\mathbf{E}_{1}^{-}\left(  z_{1}\right)   &  =\mathbf{\bar{T}}_{01}%
	\cdot\mathbf{E}_{0}^{-}\left(  z_{1}\right)  +\mathbf{\bar{R}}_{10}%
	\cdot\mathbf{E}_{1}^{+}\left(  z_{1}\right),\label{r70}\\
	\mathbf{E}_{0}^{+}\left(  z_{1}\right)   &  =\mathbf{\bar{R}}_{01}%
	\cdot\mathbf{E}_{0}^{-}\left(  z_{1}\right)  +\mathbf{\bar{T}}_{10}%
	\cdot\mathbf{E}_{1}^{+}\left(  z_{1}\right),\label{r71}\\
	\mathbf{E}_{1}^{+}\left(  z_{2}\right)   &  =\mathbf{\bar{R}}_{12}%
	\cdot\mathbf{E}_{1}^{-}\left(  z_{2}\right),\label{r72}\\
	\mathbf{E}_{2}^{-}\left(  z_{2}\right)   &  =\mathbf{\bar{T}}_{12}%
	\cdot\mathbf{E}_{1}^{-}\left(  z_{2}\right), \label{r73}%
\end{align}
where the electric field associated with a plane wave propagating a distance,
$h=\left\vert z_{2}-z_{1}\right\vert $, along the $\pm z$ direction within the
gyrotropic slab, is given by%
\begin{align}
	\mathbf{E}_{1}^{-}\left(  z_{2}\right)   &  =\mathbf{\bar{P}}_{E}^{-}%
	\cdot\mathbf{E}_{1}^{-}\left(  z_{1}\right),\label{r74}\\
	\mathbf{E}_{1}^{+}\left(  z_{1}\right)   &  =\mathbf{\bar{P}}_{E}^{+}%
	\cdot\mathbf{E}_{1}^{+}\left(  z_{2}\right), \label{r75}%
\end{align}
where $\mathbf{\bar{P}}_{E}^{m}$ denotes the spacial propagator, which
effectively propagates the electric field a distance $h$ through the slab
and takes the form
\begin{equation}
	\mathbf{\bar{P}}_{E}^{m}=\mathbf{\bar{U}}_{m}\cdot\mathbf{\bar{P}}^{m}%
	\cdot\mathbf{\bar{U}}_{m}^{-1}, \label{r76}%
\end{equation}
where
\begin{align}
	\mathbf{\bar{U}}_{m} &  =\left(
	\begin{array}
	[c]{cc}%
	\beta_{1}^{m}/k_{t,1} & \beta_{2}^{m}/k_{t,2}\\
	k_{y}\theta_{1}/k_{t,1} & k_{y}\theta_{2}/k_{t,2}%
	\end{array}
	\right)  ,\label{r77}\\
	\mathbf{\bar{P}}^{m} &  =\left(
	\begin{array}
	[c]{cc}%
	e^{jk_{z,1}h} & 0\\
	0 & e^{jk_{z,2}h}%
	\end{array}
	\right)  .\label{r78}%
\end{align}
Using (\ref{r74})-(\ref{r75}) in (\ref{r70})-(\ref{r73}) leads to
\begin{align}
	\mathbf{E}_{0}^{+}\left(  z_{1}\right)   &  =\mathbf{\bar{R}}\cdot
	\mathbf{E}_{0}^{-}\left(  z_{1}\right)  ,\label{r79}\\
	\mathbf{E}_{2}^{-}\left(  z_{2}\right)   &  =\mathbf{\bar{T}\cdot E}_{0}%
	^{-}\left(  z_{1}\right),\label{r80}%
\end{align}
where
\begin{align}
	\mathbf{\bar{R}}  &  =\mathbf{\bar{R}}_{01}+\mathbf{\bar{T}}_{10}%
	\cdot\mathbf{\bar{R}}_{12}^{\prime}\cdot\left(  \mathbf{\bar{I}}_{s}%
	-\mathbf{\bar{R}}_{10}\cdot\mathbf{\bar{R}}_{12}^{\prime}\right)  ^{-1}%
	\cdot\mathbf{\bar{T}}_{01},\label{r81}\\
	\mathbf{\bar{T}}  &  =\mathbf{\bar{T}}_{12}\cdot\mathbf{\bar{P}}_{E}^{-}%
	\cdot\left(  \mathbf{\bar{I}}_{s}-\mathbf{\bar{R}}_{10}\cdot\mathbf{\bar{R}}%
	_{12}^{\prime}\right)  ^{-1}\cdot\mathbf{\bar{T}}_{01}, \label{r82}%
\end{align}
such that $\mathbf{\bar{R}}_{12}^{\prime}=\mathbf{\bar{P}}_{E}^{+}\cdot\mathbf{\bar{R}}_{12}\cdot\mathbf{\bar{P}}_{E}^{-}$. After some algebra, we find that (\ref{r67}), (\ref{r76}), (\ref{r81}), and (\ref{r82}) may be written in numerator/denominator form as
\begin{align}
	\mathbf{\bar{R}}_{nn^{\prime}} &  =\frac{1}{\Omega^{nn^{\prime}}}\left(
	\begin{array}
	[c]{cc}%
	\Pi_{11}^{nn^{\prime}} & \Pi_{12}^{nn^{\prime}}/k_{y}\\
	k_{y}\Pi_{21}^{nn^{\prime}} & \Pi_{22}^{nn^{\prime}}%
	\end{array}
	\right)  ,\label{r84}\\
	\mathbf{\bar{P}}_{E}^{m} &  =\frac{1}{\chi^{m}}\left(
	\begin{array}
	[c]{cc}%
	\Delta_{11}^{m} & \Delta_{12}^{m}/k_{y}\\
	k_{y}\Delta_{21} & \Delta_{22}^{m}%
	\end{array}
	\right)  ,\label{r85}\\
	\mathbf{\bar{R}} &  =\frac{1}{\Lambda\Omega^{01}}\left(
	\begin{array}
	[c]{cc}%
	\Xi_{11} & \Xi_{12}/k_{y}\\
	k_{y}\Xi_{21} & \Xi_{22}%
	\end{array}
	\right)  ,\label{rr86}\\
	\mathbf{\bar{T}} &  =\frac{\Omega^{10}\chi^{+}}{\Lambda\Omega^{01}}\left(
	\begin{array}
	[c]{cc}%
	\Psi_{11} & \Psi_{12}/k_{y}\\
	k_{y}\Psi_{21} & \Psi_{22}%
	\end{array}
	\right)  ,\label{rr87}%
\end{align}
where we define
\begin{widetext}
	\begin{align}
		\Lambda &  =\left(  \Omega^{10}\Phi-\Theta_{11}\right)  \left(
		\Omega^{10}\Phi-\Theta_{22}\right)  -\Theta_{12}\Theta_{21},\label{58}\\
		\Xi_{11} &  =\Lambda\Pi_{11}^{01}+\left(  \Omega^{10}+\Pi_{11}%
		^{10}\right)  \left(  \Upsilon_{11}\Sigma_{11}+\Upsilon_{12}\Sigma
		_{21}\right)  +\Pi_{12}^{10}\left(  \Upsilon_{21}\Sigma_{11}+\Upsilon
		_{22}\Sigma_{21}\right)  ,\label{59}\\
		\Xi_{12} &  =\Lambda\Pi_{12}^{01}+\left(  \Omega^{10}+\Pi_{11}%
		^{10}\right)  \left(  \Upsilon_{11}\Sigma_{12}+\Upsilon_{12}\Sigma
		_{22}\right)  +\Pi_{12}^{10}\left(  \Upsilon_{21}\Sigma_{12}+\Upsilon
		_{22}\Sigma_{22}\right)  ,\label{60}\\
		\Xi_{21} &  =\Lambda\Pi_{21}^{01}+\left(  \Omega^{10}+\Pi_{22}%
		^{10}\right)  \left(  \Upsilon_{21}\Sigma_{11}+\Upsilon_{22}\Sigma
		_{21}\right)  +\Pi_{21}^{10}\left(  \Upsilon_{11}\Sigma_{11}+\Upsilon
		_{12}\Sigma_{21}\right)  ,\label{61}\\
		\Xi_{22} &  =\Lambda\Pi_{22}^{01}+\left(  \Omega^{10}+\Pi_{22}%
		^{10}\right)  \left(  \Upsilon_{21}\Sigma_{12}+\Upsilon_{22}\Sigma
		_{22}\right)  +\Pi_{21}^{10}\left(  \Upsilon_{11}\Sigma_{12}+\Upsilon
		_{12}\Sigma_{22}\right)  ,\label{62}\\
		\Psi_{11} &  =\left(  \Omega^{12}+\Pi_{11}^{12}\right)  \left(
		\Delta_{11}^{-}\Sigma_{11}+\Delta_{12}^{-}\Sigma_{21}\right)  +\Pi_{12}%
		^{12}\left(  \Delta_{21}\Sigma_{11}+\Delta_{22}^{-}\Sigma_{21}\right)
		,\label{63}\\
		\Psi_{12} &  =\left(  \Omega^{12}+\Pi_{11}^{12}\right)  \left(
		\Delta_{11}^{-}\Sigma_{12}+\Delta_{12}^{-}\Sigma_{22}\right)  +\Pi_{12}%
		^{12}\left(  \Delta_{21}\Sigma_{12}+\Delta_{22}^{-}\Sigma_{22}\right)
		,\label{64}\\
		\Psi_{21} &  =\left(  \Omega^{12}+\Pi_{22}^{12}\right)  \left(
		\Delta_{21}\Sigma_{11}+\Delta_{22}^{-}\Sigma_{21}\right)  +\Pi_{21}%
		^{12}\left(  \Delta_{11}^{-}\Sigma_{11}+\Delta_{12}^{-}\Sigma_{21}\right)
		,\label{65}\\
		\Psi_{22} &  =\left(  \Omega^{12}+\Pi_{22}^{12}\right)  \left(
		\Delta_{21}\Sigma_{12}+\Delta_{22}^{-}\Sigma_{22}\right)  +\Pi_{21}%
		^{12}\left(  \Delta_{11}^{-}\Sigma_{12}+\Delta_{12}^{-}\Sigma_{22}\right)
		,\label{66}\\
		\Omega^{nn^{\prime}} &  = m_{1}m_{3}k_{z}\chi^{m_{3}}\left(  n_{E}%
		^{m_{2}}-\varepsilon_{r,0}\chi^{m_{2}}\right)  \nonumber\\
		&  +jm_{3}\chi^{m_{3}}\left[  \left(  k_{y}^{2}+k_{z}^{2}\right)  n_{A}%
		-k_{x}n_{B}^{m_{2}}+k_{x}k_{y}^{2}n_{C}^{m_{2}}-\left(  k_{x}^{2}+k_{z}%
		^{2}\right)  n_{D}^{m_{2}}\right]  ,\label{67}\\
		\Pi_{11}^{nn^{\prime}} &  = k_{z}\left[  \varepsilon_{r}\chi^{m_{2}}%
		\chi^{m_{3}}+m_{1}m_{3}k_{0}^{2}\left(  n_{A}n_{D}^{m_{2}}-n_{B}^{m_{2}}%
		n_{C}^{m_{3}}\right)  \right]  \nonumber\\
		&  +j\left(  m_{1}\chi^{m_{3}}\left[  \left(  k_{x}^{2}+k_{z}^{2}\right)
		n_{D}^{m_{2}}+k_{x}n_{B}^{m_{2}}\right]  -m_{3}\chi^{m_{2}}\left[  \left(
		k_{y}^{2}+k_{z}^{2}\right)  n_{A}+k_{x}k_{y}^{2}n_{C}^{m_{3}}\right]  \right)
		,\label{68}\\
		\Pi_{12}^{nn^{\prime}} &  = m_{1}m_{3}k_{z}k_{0}^{2}\left(  n_{D}^{m_{2}%
		}n_{B}^{m_{3}}-n_{D}^{m_{3}}n_{B}^{m_{2}}\right)  \nonumber\\
		&  +j\left[  k_{x}k_{y}^{2}\left(  m_{1}n_{D}^{m_{2}}\chi^{m_{3}}-m_{3}%
		n_{D}^{m_{3}}\chi^{m_{2}}\right)  +\left(  k_{y}^{2}+k_{z}^{2}\right)  \left(
		m_{1}n_{B}^{m_{2}}\chi^{m_{3}}-m_{3}n_{B}^{m_{3}}\chi^{m_{2}}\right)  \right]
		,\label{69}\\
		\Pi_{21}^{nn^{\prime}} &  = m_{1}m_{3}k_{z}k_{0}^{2}n_{A}\left(
		n_{C}^{m_{3}}-n_{C}^{m_{2}}\right)  \nonumber\\
		&  +j\left[  k_{x}n_{A}\left(  m_{3}\chi^{m_{2}}-m_{1}\chi^{m_{3}}\right)
		+\left(  k_{x}^{2}+k_{z}^{2}\right)  \left(  m_{3}n_{C}^{m_{3}}\chi^{m_{2}%
		}-m_{1}n_{C}^{m_{2}}\chi^{m_{3}}\right)  \right]  ,\label{70}\\
		\Pi_{22}^{nn^{\prime}} &  = k_{z}\left[  \varepsilon_{r,0}\chi^{m_{2}%
		}\chi^{m_{3}}+m_{1}m_{3}k_{0}^{2}\left(  n_{A}n_{D}^{m_{3}}-n_{C}^{m_{2}}%
		n_{B}^{m_{3}}\right)  \right]  \nonumber\\
		&  +j\left(  m_{3}\left[  k_{x}n_{B}^{m_{3}}\chi^{m_{2}}+\left(  k_{x}%
		^{2}+k_{z}^{2}\right)  n_{D}^{m_{3}}\chi^{m_{2}}\right]  -m_{1}\chi^{m_{3}%
		}\left[  k_{x}k_{y}^{2}n_{C}^{m_{2}}+\left(  k_{y}^{2}+k_{z}^{2}\right)
		n_{A}\right]  \right)  ,\label{71}%
	\end{align}
\end{widetext}
such that
\begin{align}
	\Phi &  =\Omega^{12}\chi^{+}\chi^{-},\label{72}\\
	\Upsilon_{11} &  =\Delta_{11}^{+}\left(  \Pi_{11}^{12}\Delta_{11}^{-}+\Pi
	_{12}^{12}\Delta_{21}\right)  \nonumber\\
	&  +\Delta_{12}^{+}\left(  \Pi_{21}^{12}\Delta_{11}^{-}+\Pi_{22}^{12}%
	\Delta_{21}\right)  ,\label{73}\\
	\Upsilon_{12} &  =\Delta_{11}^{+}\left(  \Pi_{11}^{12}\Delta_{12}^{-}+\Pi
	_{12}^{12}\Delta_{22}^{-}\right)  \nonumber\\
	&  +\Delta_{12}^{+}\left(  \Pi_{21}^{12}\Delta_{12}^{-}+\Pi_{22}^{12}%
	\Delta_{22}^{-}\right)  ,\label{74}\\
	\Upsilon_{21} &  =\Delta_{21}\left(  \Pi_{11}^{12}\Delta_{11}^{-}+\Pi
	_{12}^{12}\Delta_{21}\right)  \nonumber\\
	&  +\Delta_{22}^{+}\left(  \Pi_{21}^{12}\Delta_{11}^{-}+\Pi_{22}^{12}%
	\Delta_{21}\right)  ,\label{75}\\
	\Upsilon_{22} &  =\Delta_{21}\left(  \Pi_{11}^{12}\Delta_{12}^{-}+\Pi
	_{12}^{12}\Delta_{22}^{-}\right)  \nonumber\\
	&  +\Delta_{22}^{+}\left(  \Pi_{21}^{12}\Delta_{12}^{-}+\Pi_{22}^{12}%
	\Delta_{22}^{-}\right)  ,\label{76}\\
	\Theta_{11} &  =\Pi_{11}^{10}\Upsilon_{11}+\Pi_{12}^{10}\Upsilon
	_{21},\label{77}\\
	\Theta_{12} &  =\Pi_{11}^{10}\Upsilon_{12}+\Pi_{12}^{10}\Upsilon
	_{22},\label{78}\\
	\Theta_{21} &  =\Pi_{21}^{10}\Upsilon_{11}+\Pi_{22}^{10}\Upsilon
	_{21},\label{79}\\
	\Theta_{22} &  =\Pi_{21}^{10}\Upsilon_{12}+\Pi_{22}^{10}\Upsilon
	_{22},\label{80}\\
	\Sigma_{11} &  =\left(  \Omega^{10}\Phi-\Theta_{22}\right)  \left(
	\Omega^{01}+\Pi_{11}^{01}\right)  +\Theta_{12}\Pi_{21}^{01},\label{81}\\
	\Sigma_{12} &  =\left(  \Omega^{10}\Phi-\Theta_{22}\right)  \Pi_{12}%
	^{01}+\Theta_{12}\left(  \Omega^{01}+\Pi_{22}^{01}\right)  ,\label{82}\\
	\Sigma_{21} &  =\left(  \Omega^{10}\Phi-\Theta_{11}\right)  \Pi_{21}%
	^{01}+\Theta_{21}\left(  \Omega^{01}+\Pi_{11}^{01}\right)  ,\label{83}\\
	\Sigma_{22} &  =\left(  \Omega^{10}\Phi-\Theta_{11}\right)  \left(
	\Omega^{01}+\Pi_{22}^{01}\right)  +\Theta_{21}\Pi_{12}^{01},\label{84}%
\end{align}
and
\begin{align}
	n_{A} &  =\varepsilon_{g}k_{t,1}^{2}k_{t,2}^{2}\left(  \varpi_{1}\xi
	_{2}-\varpi_{2}\xi_{1}\right)  ,\label{85}\\
	n_{B}^{m} &  =\varepsilon_{g}\varpi_{1}\varpi_{2}\left(  k_{t,1}^{2}\alpha
	_{2}^{m}-k_{t,2}^{2}\alpha_{1}^{m}\right)  ,\label{86}\\
	n_{C}^{m} &  =k_{t,1}^{2}\zeta_{2}^{m}\xi_{1}-k_{t,2}^{2}\zeta_{1}^{m}\xi
	_{2},\label{87}\\
	n_{D}^{m} &  =\zeta_{2}^{m}\alpha_{1}^{m}\varpi_{1}-\zeta_{1}^{m}\alpha
	_{2}^{m}\varpi_{2},\label{88}\\
	n_{E}^{m} &  =\varepsilon_{g}k_{0}^{2}\left(  k_{t,1}^{2}\zeta_{2}^{m}%
	\varpi_{1}-k_{t,2}^{2}\zeta_{1}^{m}\varpi_{2}\right)  ,\label{89}\\
	\zeta_{i}^{m} &  =\varepsilon_{g}k_{x}\varpi_{i}-j\varepsilon_{a}\xi
	_{i}mk_{z,i},\label{90}\\
	\alpha_{i}^{m} &  =k_{x}\xi_{i}-j\varepsilon_{g}k_{0}^{2}mk_{z,i},\label{91}\\
	\chi^{m} &  =k_{t,1}^{2}\varpi_{2}\xi_{1}\alpha_{2}^{m}-k_{t,2}^{2}\varpi
	_{1}\xi_{2}\alpha_{1}^{m},\label{92}\\
	\Delta_{11} &  = k_{t,1}^{2}\varpi_{2}\xi_{1}\alpha_{2}^{m}e^{jk_{z,2}%
	h}\nonumber\\
	&  -k_{t,2}^{2}\varpi_{1}\xi_{2}\alpha_{1}^{m}e^{jk_{z,1}h},\label{93}\\
	\Delta_{12} &  =\varpi_{1}\varpi_{2}\alpha_{1}^{m}\alpha_{2}^{m}\left(
	e^{jk_{z,2}h}-e^{jk_{z,1}h}\right)  ,\label{94}\\
	\Delta_{21} &  = k_{t,1}^{2}k_{t,2}^{2}\xi_{1}\xi_{2}\left(
	e^{jk_{z,1}h}-e^{jk_{z,2}h}\right)  ,\label{95}\\
	\Delta_{22} &  = k_{t,1}^{2}\varpi_{2}\xi_{1}\alpha_{2}^{m}e^{jk_{z,1}%
	h}\nonumber\\
	&  -k_{t,2}^{2}\varpi_{1}\xi_{2}\alpha_{1}^{m}e^{jk_{z,2}h}.\label{96}%
\end{align}

\section*{}


\begin{thebibliography}{99}
\bibitem{18-ozawa2018topological} T. Ozawa, H.M Price, A. Amo,  N. Goldman, M. Hafezi, L.Lu, M. Rechtsman, D. Schuster, J. Simon, and O. Zilberberg, \emph{Topological photonics}, arXiv preprint arXiv:1802.04173 (2018).

\bibitem{19-Soljacic2014} L.Lu, J.D. Joannopoulos, and M. Solja\v{c}i\'{c},  \emph{Topological Photonics}, Nat. Photonics {\bf{8}}, 821 – 829 (2014).

\bibitem{23-hasan2010colloquium} M.Z. Hasan, and Ch L. Kane, \emph{Colloquium: topological insulators}, Reviews of Modern Physics {\bf{82}}, 3045 (2010).

\bibitem{25-rechtsman2013photonic} M.C. Rechtsman, J.M. Zeuner, Y. Plotnik, Y. Lumer, D. Podolsky, F. Dreisow,  S. Nolte, M. Segev, and A. Szameit, \emph{Photonic Floquet topological insulators}, Nature {\bf{496}}, 196 (2013).

\bibitem{26-chen2014experimental} W-J. Chen, Sh.J. Jiang, X-D. Chen, B. Zhu, L. Zhou, J-W. Dong, and Ch T. Chan, \emph{Experimental realization of photonic topological insulator in a uniaxial metacrystal waveguide}, Nature communications {\bf{5}}, 5782 (2014).

\bibitem{7-PTI-Notes}  G.W. Hanson, S. Gangaraj, and A. Nemilentsau, \emph{Notes on photonic topological insulators and scattering-protected edge states-a brief introduction}, arXiv preprint arXiv:1602.02425, (2016).

\bibitem{17-wang2009observation} Zh. Wang, Y. Chong, J.D. Joannopoulos, and M. Solja{\v{c}}i{\'c}, \emph{Observation of unidirectional backscattering-immune topological electromagnetic states}, Nature {\bf{461}}, 7265 (2009).

\bibitem{casimir} S. Fuchs, F. Lindel, R.V. Krems,  G.W. Hanson, M. Antezza, and S.Y. Buhmann,\emph{ Casimir-Lifshitz force for nonreciprocal media and applications to photonic topological insulators}, Physical Review A {\bf{6}}, 062505 (2017).

\bibitem{13-Ferrite} S.A.H. Gangaraj and G.W. Hanson, \emph{Topologically protected unidirectional surface states in biased ferrites: duality and application to directional couplers}, IEEE Antennas and Wireless Propagation Letters {\bf{16}}, 449-452 (2017). 	

\bibitem{20-wang2008reflection} Zh. Wang, YD.Chong, J.D. Joannopoulos, and M. Solja{\v{c}}i{\'c}, \emph{Reflection-free one-way edge modes in a gyromagnetic photonic crystal}, Physical review letters {\bf{100}}, 013905 (2008).

\bibitem{21-yu2008one} Z. Yu, G. Veronis, Zh. Wang, and Sh. Fan, \emph{One-way electromagnetic waveguide formed at the interface between a plasmonic metal under a static magnetic field and a photonic crystal}, Physical review letters {\bf{100}}, 023902 (2008).

\bibitem{22-yang2016one} B. Yang, M. Lawrence, W.Gao,  Q. Guo, and Sh. Zhang, \emph{One-way helical electromagnetic wave propagation supported by magnetized plasma}, Scientific reports {\bf{6}}, 21461 (2016).

\bibitem{14-Mario-chern} M. Silveirinha, \emph{Chern invariants for continuous media}, Physical Review B {\bf{92}},125153 (2015).

\bibitem{15-Haldane-chern} FDM. Haldane and S. Raghu, \emph{Possible realization of directional optical waveguides in photonic crystals with broken time-reversal symmetry}, Physical review letters {\bf{100}},013904 (2008).

\bibitem{16-raghu-chern} S. Raghu and F.D. M Haldane, \emph{Analogs of quantum-Hall-effect edge states in photonic crystals}, Physical Review A {\bf{78}}, 033834 (2008).

\bibitem{27-gangaraj2017berry} S.A.H. Gangaraj, M.G. Silveirinha, and G.W. Hanson, \emph{Berry phase, Berry connection, and Chern number for a continuum bianisotropic material from a classical electromagnetics perspective}, IEEE Journal on Multiscale and Multiphysics Computational Techniques {\bf{2}}, 3-17 (2017).

\bibitem{28-skirlo2014multimode} S.A. Skirlo,  L. Lu, and M. Solja{\v{c}}i{\'c}, \emph{Multimode one-way waveguides of large Chern numbers}, Physical review letters {\bf{113}}, 113904 (2014).

\bibitem{29-khanikaev2013photonic} A.B. Khanikaev, S.H. Mousavi, W-K. Tse, M. Kargarian, A.H. MacDonald, and G. Shvets, \emph{Photonic topological insulators}, Nature materials {\bf{12}}, 233 (2013).

\bibitem{6-PRL} S.A.H. Gangaraj, G.W. Hanson, M. Antezza, and M.G Silveirinha, \emph{Spontaneous lateral atomic recoil force close to a photonic topological material}, Physical Review B {\bf{97}}, 201108 (2018).

\bibitem{9-Trully}  S. Gangaraj, G.W. Hanson, M.G. Silveirinha, K. Shastri, M. Antezza, and F. Monticone, \emph{Truly unidirectional excitation and propagation of diffractionless surface plasmon-polaritons}, arXiv preprint arXiv:1811.00463, (2018).

\bibitem{coordinateBook} H.C. Chen, \emph{Theory of electromagnetic waves: a coordinate-free approach}, McGraw-Hill New York, (1983).

\bibitem{Shi} S. Buddhiraju, Y.Shi, A.song, C.Wojcik, M.minkov, I.Williamson, A.Dutt, S.Fan, \emph{ Absence of unidirectionally propagating surface plasmon-polaritons in nonreciprocal plasmonics}, arXiv:1809.05100, (2018).

\bibitem{10-weyl} W. Gao, B. Yang, M. Lawrence, F. Fang,  B.B{\'e}ri, and S. Zhang, \emph{Photonic Weyl degeneracies in magnetized plasma}, Nature communications {\bf{7}}, 12435 (2016).


\bibitem{2-silveirinha2017topological} M.G. Silveirinha, \emph{Topological angular momentum and radiative heat transport in closed orbits}, Physical Review B {\bf{95}},115103 (2017).

\bibitem{3-PRA} M. G. Silveirinha, S. A. H Gangaraj, G.W. Hanson, and M.Antezza, \emph{Fluctuation-induced forces on an atom near a photonic topological material}, Physical Review A {\bf{97}}, 022509 (2018).

\bibitem{4-davoyan2013theory} A.R. Davoyan and N. Engheta, \emph{Theory of wave propagation in magnetized near-zero-epsilon metamaterials: evidence for one-way photonic states and magnetically switched transparency and opacity}, Physical review letters {\bf{111}}, 257401 (2013).

\bibitem{5-PRA-june} S.A.H. Gangaraj, G.W. Hanson, and M. Antezza, \emph{Robust entanglement with three-dimensional nonreciprocal photonic topological insulators}, Physical Review A {\bf{95}}, 063807 (2017).

\bibitem{8-three-defect}  S.A.H Gangaraj, A. Nemilentsau, and G.W. Hanson \emph{The effects of three-dimensional defects on one-way surface plasmon propagation for photonic topological insulators comprised of continuum media}, Scientific reports {\bf{6}}, 30055(2016).

\bibitem{11-mario-optical} M.G. Silveirinha, \emph{Optical instabilities and spontaneous light emission by polarizable moving matter}, Physical Review X {\bf{4}}, 031013 (2014).

\end{thebibliography}
\end{document}